\documentclass{xjkas}

\def\year{2015} 
\def\month{October} 
\def\volume{00} 
\def\beginpage{1} 
\def\endpage{13} 
\def\received{August 31, 2015} 
\def\accepted{October 15, 2015} 
\date{Received \received; accepted \accepted}

\usepackage{flushend} 
\usepackage[breaklinks, colorlinks, citecolor=blue, linkcolor=blue, menucolor=blue, urlcolor=blue]{hyperref}

\def\deg{^{\circ}}

\title{
PAGaN II: The Evolution of AGN jets on Sub-Parsec Scales\thanks{Part of a special issue on the Korean VLBI Network (KVN)}
}

\author[1]{Junghwan~Oh}
\author[1]{Sascha~Trippe}
\author[2,3]{Sincheol~Kang}
\author[1,4]{Jae-Young~Kim}
\author[1]{Jong-Ho~Park}
\author[1]{Taeseok~Lee}
\author[1]{Daewon~Kim}
\author[3]{Motoki~Kino}
\author[3]{Sang-Sung~Lee}
\author[3]{Bong~Won~Sohn}

\affil[1]{Department of Physics and Astronomy, Seoul National University, Gwanak-gu, Seoul 08826, Korea \email{joh@astro.snu.ac.kr, trippe@astro.snu.ac.kr}}
\affil[2]{University of Science and Technology, Yuseong-gu, Daejeon 34113, Korea}
\affil[3]{Korea Astronomy and Space Science Institute, Yuseong-gu, Daejeon 34055, Korea}
\affil[4]{Max-Planck Institut f\"{u}r Radioastronomie, Auf dem H\"{u}gel 69, D-53121 Bonn, Germany}

\def\corrauthor{
S. Trippe
}

\def\runningauthor{
Oh et al.
}

\def\runningtitle{
PAGaN II: Evolution of AGN Jets on Sub-Parsec Scales
}

\def\keywords{
galaxies: active, jets --- radio continuum: galaxies --- techniques: interferometric
}

\def\abstracttext{
We report first results from KVN and VERA Array (KaVA) VLBI observations obtained in the frame of our \emph{P}lasma-physics of \emph{A}ctive \emph{Ga}lactic \emph{N}uclei (PAGaN) project. We observed eight selected AGN at 22 and 43 GHz in single polarization (LCP) between March 2014 and April 2015. Each source was observed for 6 to 8 hours per observing run to maximize the $uv$ coverage. We obtained a total of 15 deep high-resolution images permitting the identification of individual circular Gaussian jet components and three spectral index maps of BL Lac, 3C~111 and 3C~345 from simultaneous dual-frequency observations. The spectral index maps show trends in agreement with general expectations -- flat core and steep jets -- while the actual value of the spectral index for jets shows indications for a dependence on AGN type. We analyzed the kinematics of jet components of BL Lac and 3C~111, detecting superluminal proper motions with maximum apparent speeds of about $5c$. This constrains the lower limits of the intrinsic component velocities to $\sim0.98c$ and the upper limits of the angle between jet and line of sight to $\sim$20$\deg$. In agreement with global jet expansion, jet components show systematically larger diameters $d$ at larger core distances $r$, following the global relation $d\approx0.2r$, albeit within substantial scatter.
}

\begin{document}
\jkashead 


\section{Introduction\label{sec:intro}}

Active Galactic Nuclei (AGN) are the most powerful persistent sources of energy in the universe. It is widely accepted that their large powers, sometimes reaching up to $10^{15}$ $L_{\odot}$, are provided by accretion of interstellar matter onto supermassive black holes \citep[e.g.,][]{Ferrarese2005, Beckmann2012}. Such an extreme environment provides a natural laboratory for the study a variety of high-energy physical processes. Especially at radio frequencies, AGN tend to show strong jet outflows extending out to Megaparsec scales emitting synchrotron continuum radiation. Images from Very Long Baseline Interferometry (VLBI) studies \citep[e.g.,][]{Lister2009a,Lister2009b} have revealed that jets are highly collimated and occasionally show extreme kinematics leading to apparent superluminal proper motions of jet components. Despite of decade-long research efforts, there is no established mechanism providing a full explanation for the formation and launching of jets. Although already \citet{Blandford1977} provided a theoretical model which explains the formation of jets via the interplay of accreted matter, black hole rotation, and magnetic fields in accretion disks, the details are insufficiently understood as yet.

The internal plasma-physical conditions of jets find expression in (1) their spatial structure and kinematics, (2) characteristic variations of optical depth, and (3) the strength and orientation of magnetic field. Apparent superluminal proper motions of jets can be detected by tracking bright ''knots'' (i.e., individual, well-separated components) moving away from an optically thick core. For assessing the structure and kinematics of jets, high resolution mapping is essential to distinguish and track the discrete knots. Furthermore, the observation should be conducted at two (or more) frequencies simultaneously in order to obtain a spectral index map of the source, which in turn provides information on optical depths of core and jets. Furthermore, polarimetric observations, if available, unveil the geometries and strengths of magnetic fields.

The KVN and VERA Array (KaVA), the combination of the Korean VLBI Network (KVN) and the Japanese VLBI Exploration of Radio Astrometry (VERA) radio arrays, is ideal for such studies. KVN, a recently built array with three 21-meter antennas, has unique properties such as four-channel dual-polarization receivers operating simultaneously at frequencies of 22, 43, 86 and 129 GHz \citep{Lee2014}. The small number of baselines, however, limits its imaging capability as well as relatively short (up to 500 km) baseline length. On the other hand, VERA has characteristics that are complementary to those of KVN in many ways. It has longer baselines (up to 2300 km) and four antennas but lacks shorter baselines. The joint array, KaVA, employs a total of seven antennas at frequencies of 22 and 43 GHz, resulting in angular resolutions of 1.2 and 0.6 mas, respectively. Although simultaneous multi-frequency observation is not possible, this limit can be overcome by observing targets consecutively at both frequencies. The imaging capability of KaVA has been discussed in detail by \citet{Niinuma2014}.

Motivated by the arguments given above, we commenced the \emph{P}lasma-physics of \emph{A}ctive \emph{Ga}lactic \emph{N}uclei (PAGaN) project which studies the plasma-physical properties of AGN jets via dedicated multi-frequency and polarimetric (where possible) VLBI observations. In the present paper (PAGaN~II), we present and discuss our first results from KaVA mapping observations. In a companion paper (PAGaN~I, \citealt{Kim2015}), we present first results from polarimetric observations with KVN.

\section{Observations and Data Reduction\label{sec:obs}}

\begin{table}[t!]
\caption{PAGaN observations with KaVA in 2014 and 2015 \label{tab:table1}}
\centering
\begin{tabular}{lllrr}
\toprule
Year         & Date    & Source & Frequency & Obs. time  \\
             &         &        & (GHz)     & (hrs)      \\
\midrule
2014         & Apr. 17 & BL LAC   & 43 & 6    \\
2014         & Mar. 15 & 3C111    & 43 & 6  \\ \addlinespace
2014         & Oct. 22 & BL LAC   & 22 & 7  \\
2014         & Oct. 22 & 1624+690 & 22  & 7  \\
2014         & Oct. 23 & 3C111    & 22  & 8  \\
2014         & Nov. 3  & 1055+018 & 22  & 6  \\
2014         & Nov. 3  & 3C84     & 22  & 6  \\
2014         & Nov. 4  & 0133+476 & 22  & 6  \\
2014         & Nov. 5  & 3C120    & 22  & 6  \\ \addlinespace
2015$^{\rm a}$         & Mar. 31 & 3C345 & 22  & 7.5    \\
2015         & Apr. 1  & 3C345    & 43  & 7.5  \\
2015$^{\rm a}$         & Mar. 31 & BL LAC   & 22 & 8   \\
2015         & Apr. 1  & BL LAC   & 43  & 8  \\
2015$^{\rm a}$        & Apr. 1  & 3C111    & 22 & 7.5   \\
2015         & Apr. 2  & 3C111    & 43 & 7.5  \\
\bottomrule
\end{tabular}
\tabnote{
$^{\rm a}$  Data from VERA station Iriki not available.
}
\vspace{1em}
\end{table}

We initially selected seven radio-bright AGN with extended structure on milliarcsecond scales: 1055+018, 0133+476, 1642+690, 3C~120, 3C~84, 3C~111, and BL Lacertae (BL Lac). In 2015, we excluded sources with relatively high redshift (1055+018, 0133+476, and 1642+690) because the fine structure of their jets was not obvious at the relatively coarse angular scales. We also replaced 3C~84 by 3C~345, since observation data for 3C~84 can be obtained from other observations that are using 3C~84 as a calibrator. Accordingly, the total number of targets observed at least once is eight.

As summarized in Table~\ref{tab:table1}, our KaVA observations cover three seasons (spring 2014, fall 2014, spring 2015). Between two and seven sessions were carried out in each season. In 2014, 43 GHz and 22 GHz data were obtained separately in spring and fall, respectively. In spring 2015, 22 and 43 GHz observations for any given source were conducted within two consecutive days. In order to obtain good $uv$ coverage, all observations covered substantial fractions of full tracks, each consisting of multiple 10-minute scans on source. Each source was observed for 6--8 hours in 2014 and 7.5--8 hours in 2015. Single-polarization (left-handed circular, LHC) light was recorded at a rate of 1 Gbps, resulting in a bandwidth of 256 MHz. The observation mode was C5, which provides 16 intermediate frequencies (IF) and 128 channels in each IF. All seven antennas in KaVA were used except for the 22-GHz observation in 2015 when data from VERA Iriki were lost for an unknown reason after correlation. The correlation was performed in the Korea-Japan Correlation Center (KJCC) in Daejeon \citep{Yeom2009, Lee2015}.

We calibrated the correlated data using the NRAO Astronomical Image Processing System (AIPS) software package. At the beginning, we clipped channels 1--12 and 116--128 (out of 128 in total) from each IF to avoid known band-edge effects. In AIPS, we performed a-priori amplitude calibration with the task \texttt{APCAL}. This task uses the opacity-corrected system temperature ($T^{\ast}_{\rm sys}$) and gain curve of each antenna which are measured during the observations. Next, we ran the task \texttt{FRING} for fringe fitting. After calibration, we imaged our data with the Caltech Difmap software package. We deconvolved our maps with \texttt{CLEAN} and applied phase self-calibration repeatedly, and then applied amplitude self-calibration starting with long ($10^{6}$ minutes) time interval to shorter time range. Once \texttt{CLEAN}ed maps were made, we fitted sets of circular Gaussian model components to the maps using the DIFMAP function \texttt{modelfit}. We determined component locations with point-like models and then iterated \texttt{modelfit} 3--5 times to fit their sizes and fluxes \citep{Jorstad2005}. For each model component, we extracted $S_{\rm tot}$ (total flux), $S_{\rm peak}$ (peak intensity), $r$ (radial distance from the core), $\theta$ (the position angle of the component measured from north to east), and $d$ (size), with the uncertainties of the parameters calculated according to \citet{Lee2008}.

\section{Results\label{sec:result}}

\begin{table*}[t!]
\caption{Global properties of the images of our target sources\label{tab:table2}}
\centering
\begin{tabular}{lrrrrrrr}
\toprule
Source & Epoch & Frequency & Total flux & Peak            & Image rms & Beam size                  & Image scale        \\
           &           & (GHz) &  (Jy)      & (Jy/beam)    &   (mJy)      & (mas$\times$mas)   & (pc/mas)  \\
\midrule
1055+018 & 2014-11-03 & 22 & 3.86 & 3.6 & 1.9 & 1.68 $\times$ 1.29 & 7.78 \\
0133+476 & 2014-11-04 & 22 & 1.73 & 1.63 & 0.96 & 1.5 $\times$ 1.06 & 7.7 \\
1642+690 & 2014-10-22 & 22 & 0.87 & 0.61 & 0.99 & 1.34 $\times$ 1.15 & 7.35 \\
3C120 & 2014-11-05 & 22 & 1.65 & 1.45 & 1.23 & 1.44 $\times$ 1.15 & 0.65 \\
3C84 & 2014-11-03 & 22 & 21.88 & 10.1 & 17.4 & 1.43 $\times$ 1.04 & 0.35 \\
BL LAC & 2014-10-22 & 22 & 3.2 & 2.54 & 2.33 & 1.1 $\times$ 0.873 & 1.29 \\
       & 2015-03-31 & 22 & 2.01 & 1.54 & 2.71 & 1.13 $\times$ 1.04 & \\
       & 2014-04-17 & 43 & 3.43 & 2.86 & 3 & 0.648 $\times$ 0.528 & \\
       & 2015-04-01 & 43 & 1.96 & 1.45 & 0.92 & 0.638 $\times$ 0.601 & \\
3C111 & 2014-10-23 & 22 & 2.64 & 1.7 & 0.93 & 1.28 $\times$ 1.14 & 0.95 \\
      & 2015-04-01 & 22 & 1.93 & 1.17 & 1.6 & 1.24 $\times$ 1.08 & \\
      & 2014-03-15 & 43 & 1.68 & 1.06 & 1.05 & 0.666 $\times$ 0.602 & \\
      & 2015-04-02 & 43 & 1.5 & 1.01 & 1.93 & 0.613 $\times$ 0.526 \\
3C345 & 2015-03-31 & 22 & 2.76 & 1.91 & 3.7 & 1.05 $\times$ 1.01 & 6.63 \\
      & 2015-04-01 & 43 & 2.07 & 1.63 & 1.55 & 0.655 $\times$ 0.613 \\   
\bottomrule
\end{tabular}
\end{table*}

In total eight sources were observed for at least one epoch each. Four targets were observed only at 22 GHz. Data for two sources -- 3C~111 and BL Lac -- were obtained for all epochs and all frequencies. Observational results for all sources are listed in Table~\ref{tab:table2}. In the following, naturally weighted [\texttt{uvweight (0,$-1$)}] images of each source are shown. Blue circles with crosses represent model components. The minimum contour level in any given map is three times the rms noise. The labels ``C'' and ``J'' denote cores and jet components, respectively. We numbered jet components sequentially from the core outward, with ``J0'' indicating a jet component newly launched between two observing epochs. Detailed information for parameters of each source is listed in Table~\ref{tab:table12}.

\begin{figure}[t!]
\centering
\includegraphics[width=80mm]{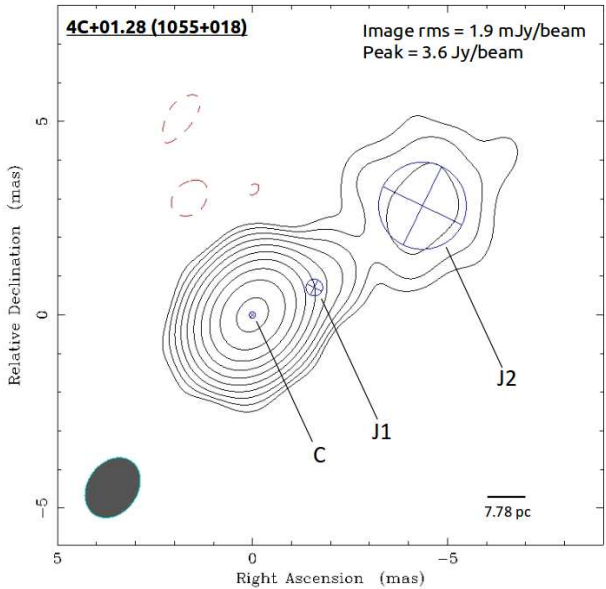}
\caption{1055+018 (4C+01.28) at 22 GHz. The synthesized beam and the angular scale are shown in the bottom-left-corner and bottom-right-corner, respectively. The image rms and peak intensity is indicated in the top-right-corner. The minimum contour level is 3 times of image rms. Blue circles with crosses are Gaussian model fitted jet components. The label ''C'' denotes the radio core, ''J'' represents jet components. Jet components are numbered in order of position from the core.\label{fig:fig1}}
\end{figure}

\begin{figure}[t!]
\centering
\includegraphics[width=80mm]{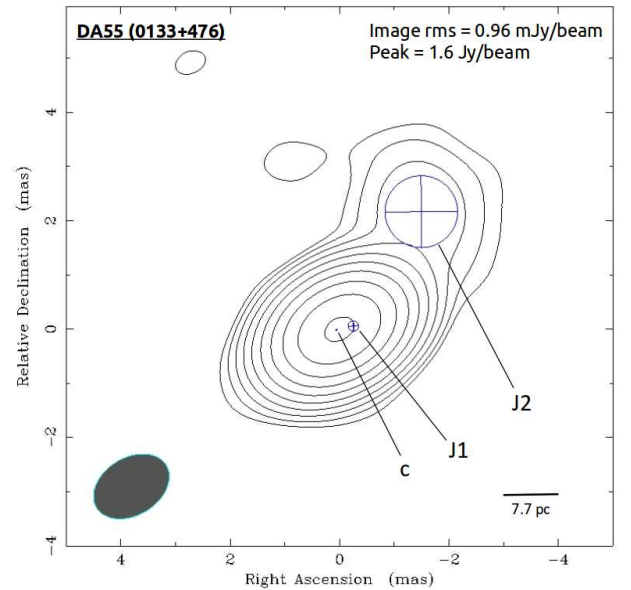}
\caption{0133+476 (DA55) at 22 GHz. Indicators are the same as in Figure~\ref{fig:fig1}.\label{fig:fig2}}
\end{figure}

\begin{figure}[t!]
\centering
\includegraphics[width=70mm]{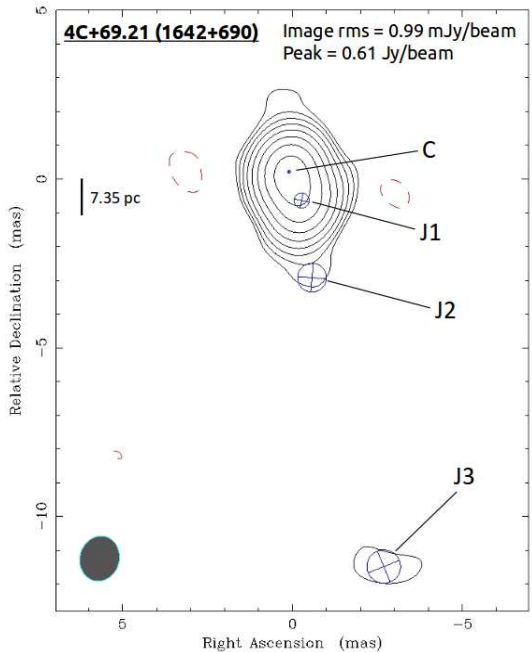}
\caption{1642+690 (4C+69.21) at 22 GHz. The angular scale is indicated on the left. Everything else is the same as in Figure~\ref{fig:fig1}.\label{fig:fig3}}
\end{figure}

\subsection{1055+018 (4C+01.28)\label{sec:1055+018}}

1055+018 (4C+01.28) is a quasar with a redshift of 0.888 \citep{Shaw2012}. The corresponding angular scale is 7.78 pc/mas. \citet{Lister2013} found a maximum apparent jet speed of $8.05c$ (with $c$ being the speed of light) based on four epochs of Very Long Baseline Array (VLBA) 15.4 GHz observations obtained within the frame of the MOJAVE program from 1998 to 2003. In our data (shown in Figure~\ref{fig:fig1}), 1055+018 shows a core-dominated structure and a bulky jet in north-western direction. The jet is extended up to $\sim$7 mas. The total flux of the entire source is 3.86 Jy with a peak intensity of 3.6 Jy/beam. Including the core, three model components can be found. Since the MOJAVE program stopped monitoring this source in October 2012, our map is one of the most recent high-resolution images of 1055+018 (the most recent observation being from December 2014 by the Boston University blazar monitoring VLBA program at 43 GHz).

\subsection{0133+476 (DA55)\label{sec:0133+476}}

Figure~\ref{fig:fig2} shows 0133+476 (DA 55), a flat-spectrum radio quasar (FSRQ) with a redshift of 0.859 (similar to 1055+018). The corresponding angular scale is 7.7 pc/mas. The source is also known as a $\gamma$-ray emitter \citep[cf.][and references therein]{Nagai2013}. \citet{Lister2013} found a maximum apparent jet speed of $15.4c$. Our map shows a jet in north-west direction extending out to $\sim$4 mas. The jet shows a rather continuous shape; this indicates that the angular resolution of KaVA at 22 GHz is not enough to resolve the detailed structure. Even though, we were able to detect three model components (including the core). The total flux of the source is 1.73 Jy, with a peak intensity of 1.63 Jy/beam. Like for 1055+018, monitoring on this source by the MOJAVE program stopped in July 2013.

\subsection{1642+690 (4C+69.21)\label{sec:1642+690}}

1642+690 (4C+69.21), shown in Figure~\ref{fig:fig3}, is a radio loud quasar with a redshift of 0.751 \citep{Lawrence1986}. The corresponding angular scale is 7.35 pc/mas. Several works reported that the source shows apparent superluminal jet with a speed of $9c$ \citep{Venturi1997, Lister2013}. In our data, we only found a core-dominated morphology with a very faint jet component south of the core, approximately $\sim$12 mas from the core. Even though the detailed source structure is not resolved, we found four model components that are expanding along the jet. 1642+690 has a total flux of 0.87 Jy with a peak intensity of 0.61 Jy/beam, making it the faintest among our targets. The MOJAVE program stopped observations of this source in September 2012.

\subsection{3C~120\label{sec:3c120}}

\begin{figure}[t!]
\centering
\includegraphics[width=80mm]{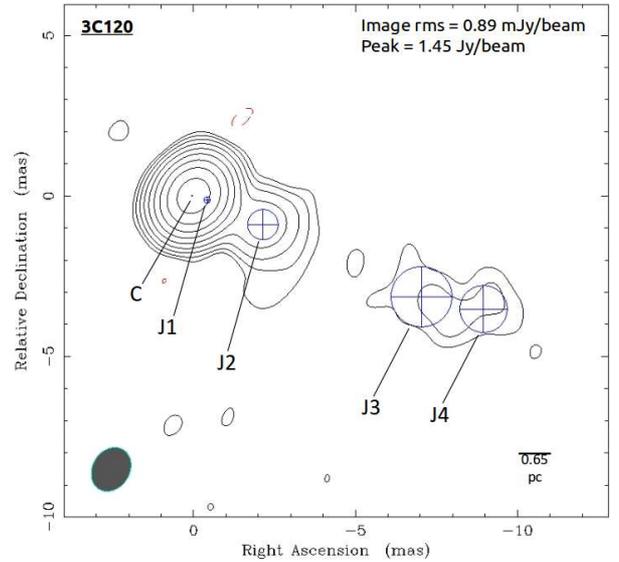}
\caption{3C~120 at 22 GHz. Indicators are the same as in Figure~\ref{fig:fig1}.\label{fig:fig4}}
\end{figure}

3C~120 (Figure~\ref{fig:fig4}) is well known for its extended structure toward the south-west. The source is classified as radio galaxy with a redshift of 0.033 \citep{Michel1988}, corresponding to an angular scale of 0.65 pc/mas. Our map shows an extended jet reaching out to more than 10 mas from the core. 3C~120 has been studied with respect to its $\gamma$-ray emission \citep[e.g.,][]{Sahakyan2015, Casadio2015} and its polarization \citep[e.g.,][and references therein]{Gomez2008}. Here, we focus on its jet kinematics. The known maximum apparent jet speed is $6.43c$ \citep{Lister2013}. With additional data from future KaVA observation, we expect to trace the motion of the jet components and to test its kinematics. We were able to detect five model components, which also shows a general trend of increasing size as function of distance from the core. The total flux of the source is 1.65 Jy, with a peak intensity of 1.45 Jy/beam.

\subsection{3C~84\label{sec:3c84}}

\begin{figure}[t!]
\centering
\includegraphics[width=80mm]{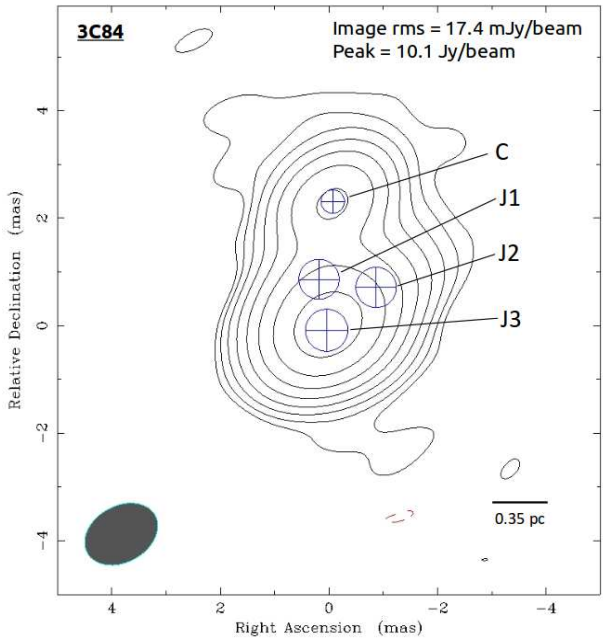}
\caption{3C~84 at 22 GHz. Indicators are the same as in Figure~\ref{fig:fig1}. The map center is placed onto the brightest component J3, not the core. \label{fig:fig5}}
\end{figure}

3C~84 (Figure~\ref{fig:fig5}) might well be one of the most ``popular'' radio galaxies in radio astronomy since it greatly increased its activity starting in 2005 \citep{Abdo2009}, and is the brightest source among our targets (21.88 Jy in total flux, 10.1 Jy/beam in peak intensity). As already noted by previous studies \citep[e.g.,][and references therein]{Nagai2014}, the jet component J3, which is located $\sim$2 mas south of the core, is brighter than the core itself (causing the phase center of our map to be placed onto J3). 
Given its brightness, we used 3C~84 as amplitude calibrator in our observations whenever it was available. It is well known that 3C~84 is effectively unpolarized, probably due to strong Faraday depolarization \citep{Walker2000, Trippe2012} (accordingly, 3C~84 was used as a polarimetric calibrator in PAGaN I). Since it keeps being observed as a calibrator for other targets, we excluded 3C~84 from dedicated observations in 2015. 
3C~84 has a redshift of 0.0176 \citep{Strauss1992} and a corresponding angular resolution of 0.35 pc/mas, making it the closest of our targets.

\subsection{BL Lac\label{sec:bllac}}

\begin{figure*}[t!]
\centering
\includegraphics[width=70mm]{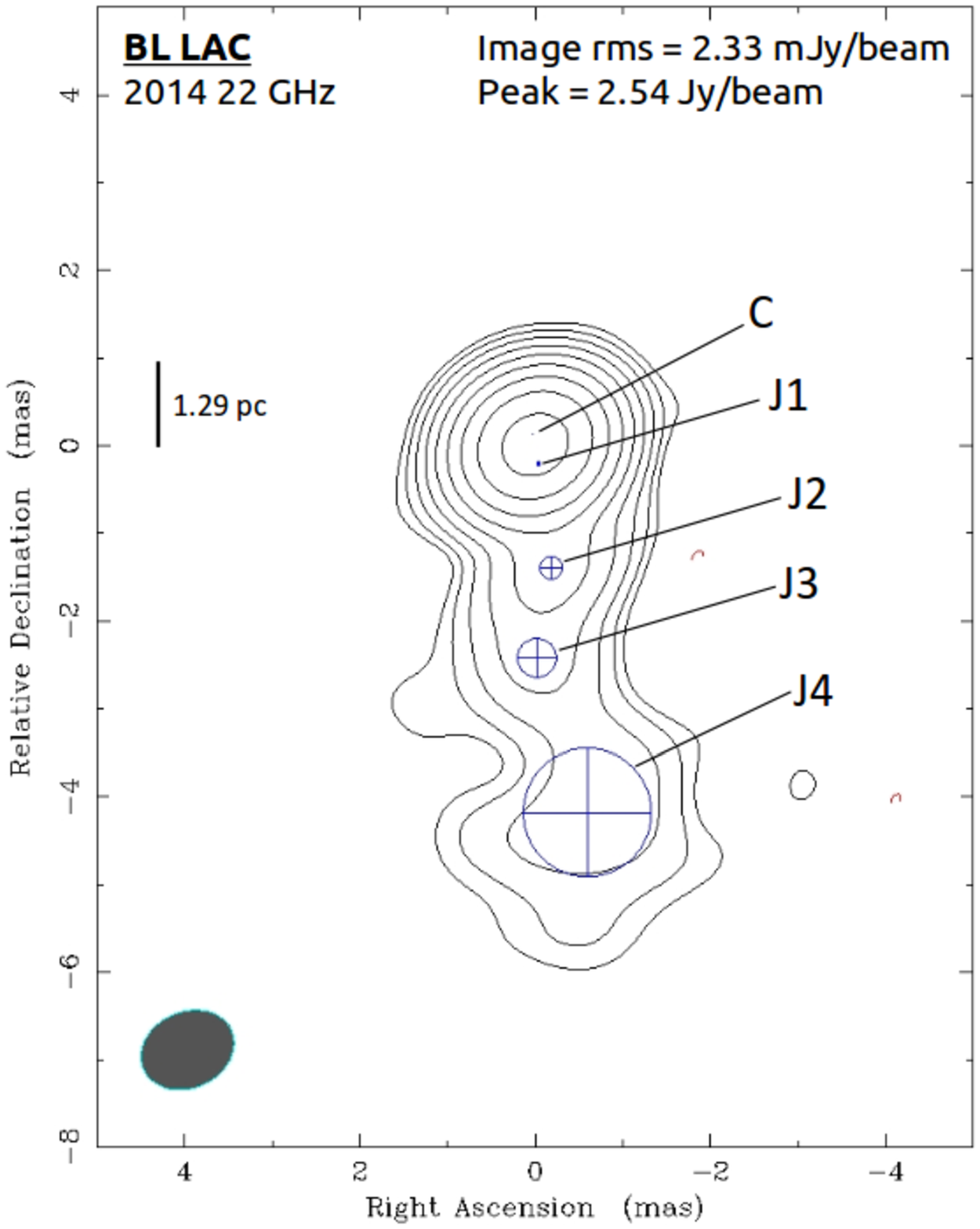}\hskip7mm
\includegraphics[width=70mm]{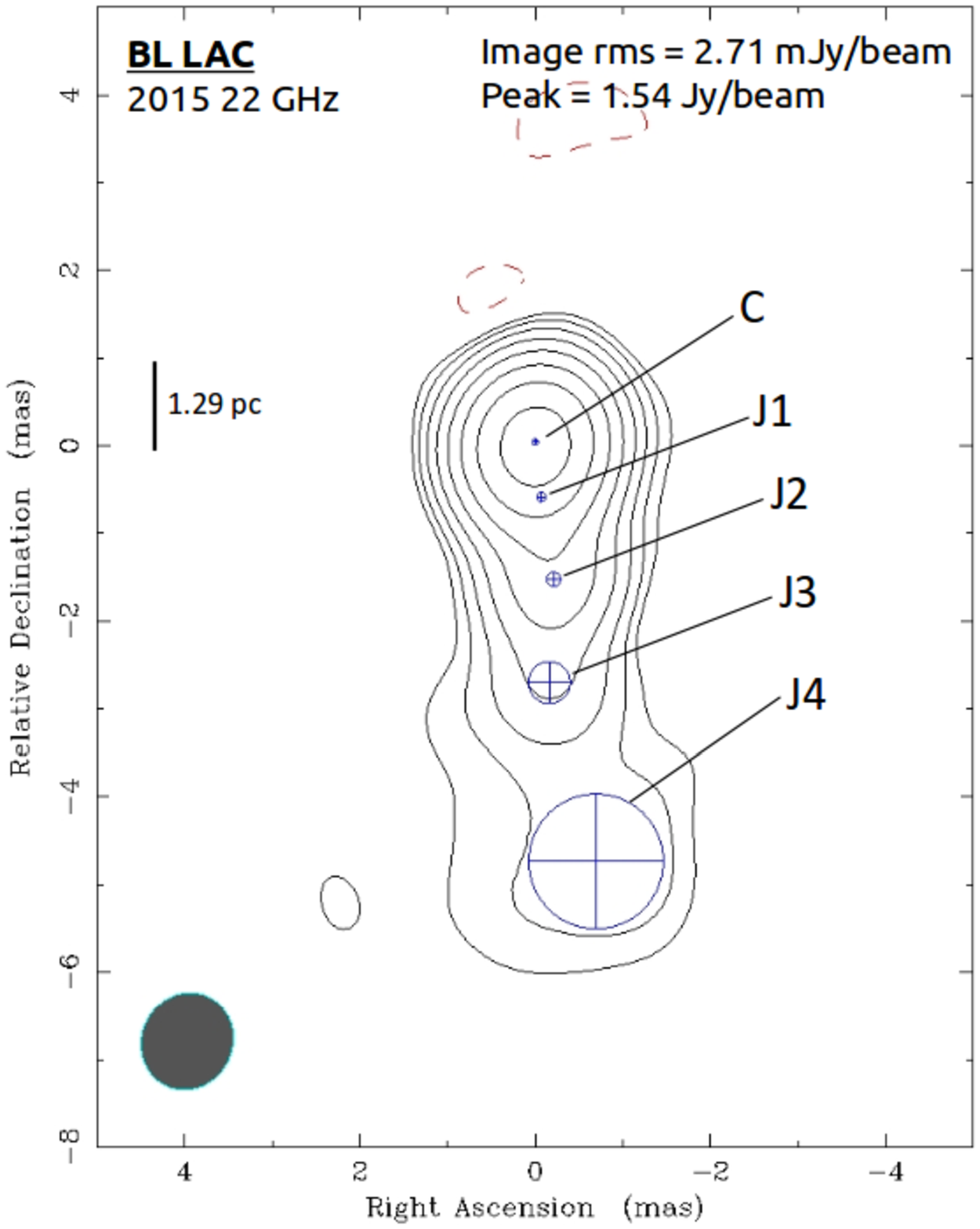} \\ 
\includegraphics[width=70mm]{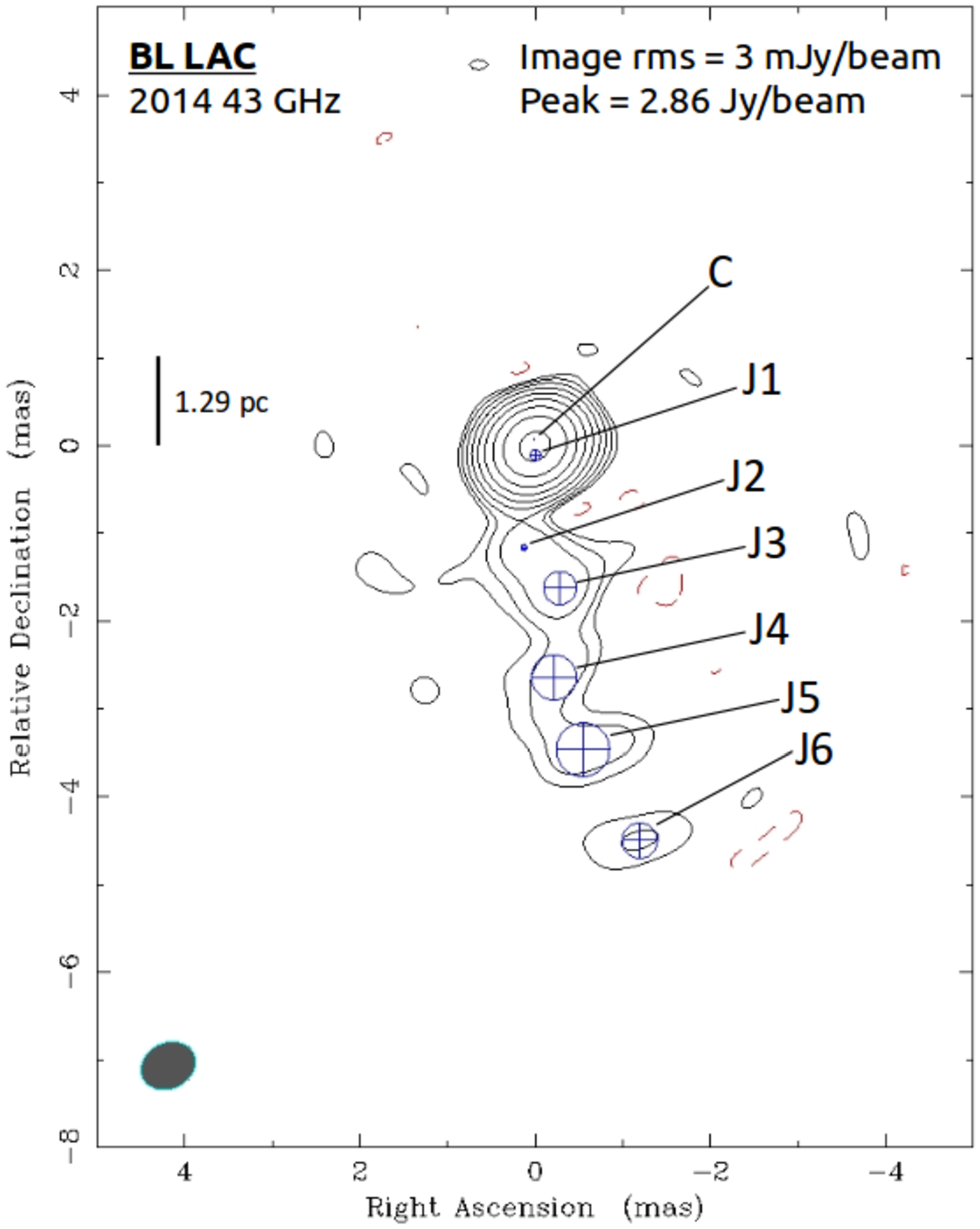}\hskip7mm
\includegraphics[width=70mm]{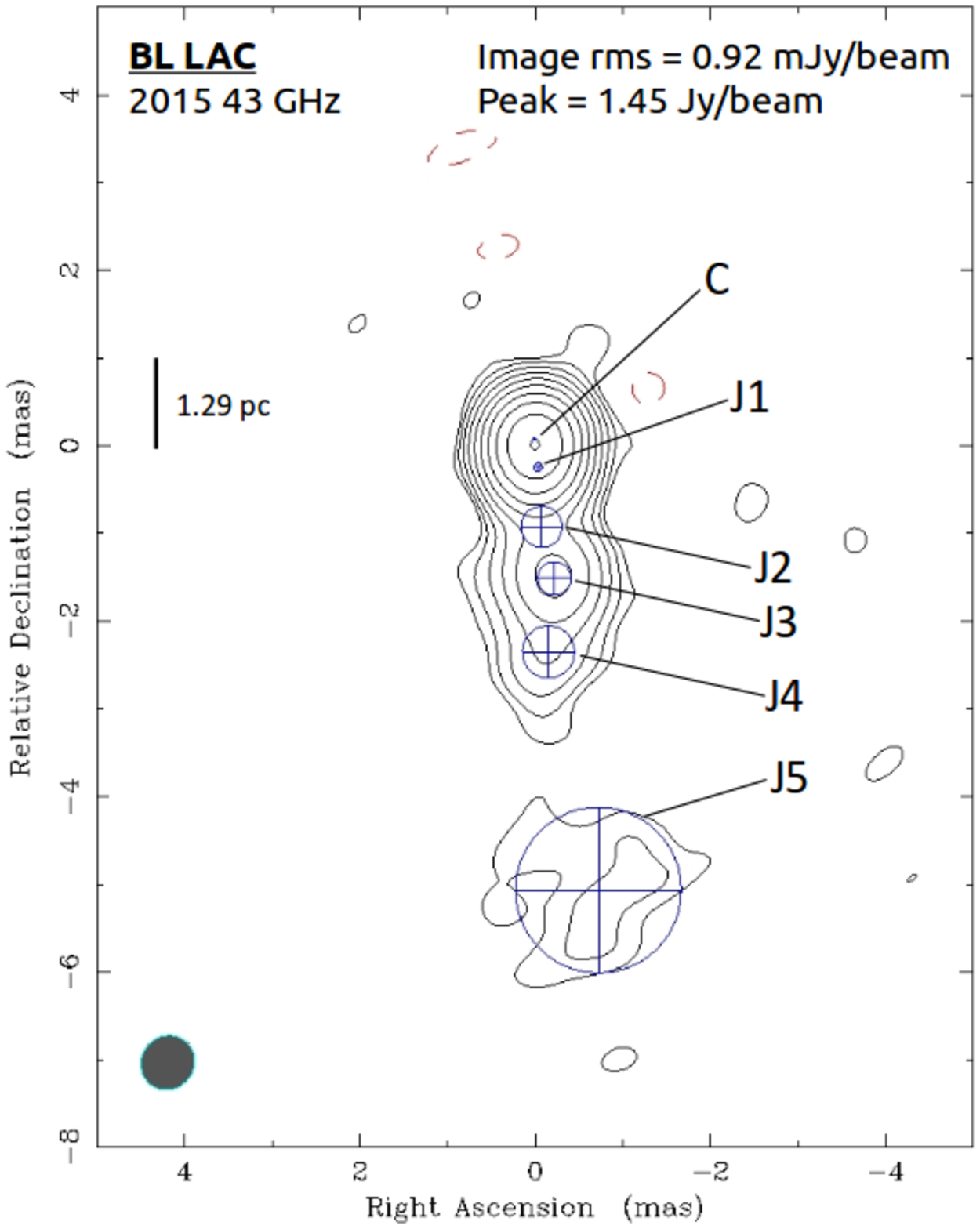}
\caption{Maps of BL Lacertae. \emph{Top left:} At 22 GHz in October 2014. \emph{Top right:} At 22 GHz in March 2015. \emph{Bottom left:} At 43 GHz in April 2014. \emph{Bottom right:} At 43 GHz in April 2015. Indicators are the same as in Figure~\ref{fig:fig3}.\label{fig:fig6}}
\end{figure*}

BL Lac (2200+420) has a redshift of 0.0686 \citep{Vermeulen1995}, which corresponds to an angular scale of 1.29 pc/mas. We were able to obtain four deep high resolution images (Figure~\ref{fig:fig6}) at 22 and 43 GHz in 2014 and 2015. In 2015, quasi-simultaneous dual-frequency observations were possible. The source is characterized by its extended and complicated jet structure toward the south. BL Lac shows significant changes in the total flux and map peak within about one year, as well as in the morphology of the jet. The total flux was 3.2 Jy in 2014 and 2.01 Jy in 2015 at 22 GHz, and 3.43 Jy and 1.96 Jy at 43 GHz. We found 5--7 jet components in each map.

\subsection{3C~111\label{sec:3c111}}

\begin{figure*}[t!]
\centering
\includegraphics[width=80mm]{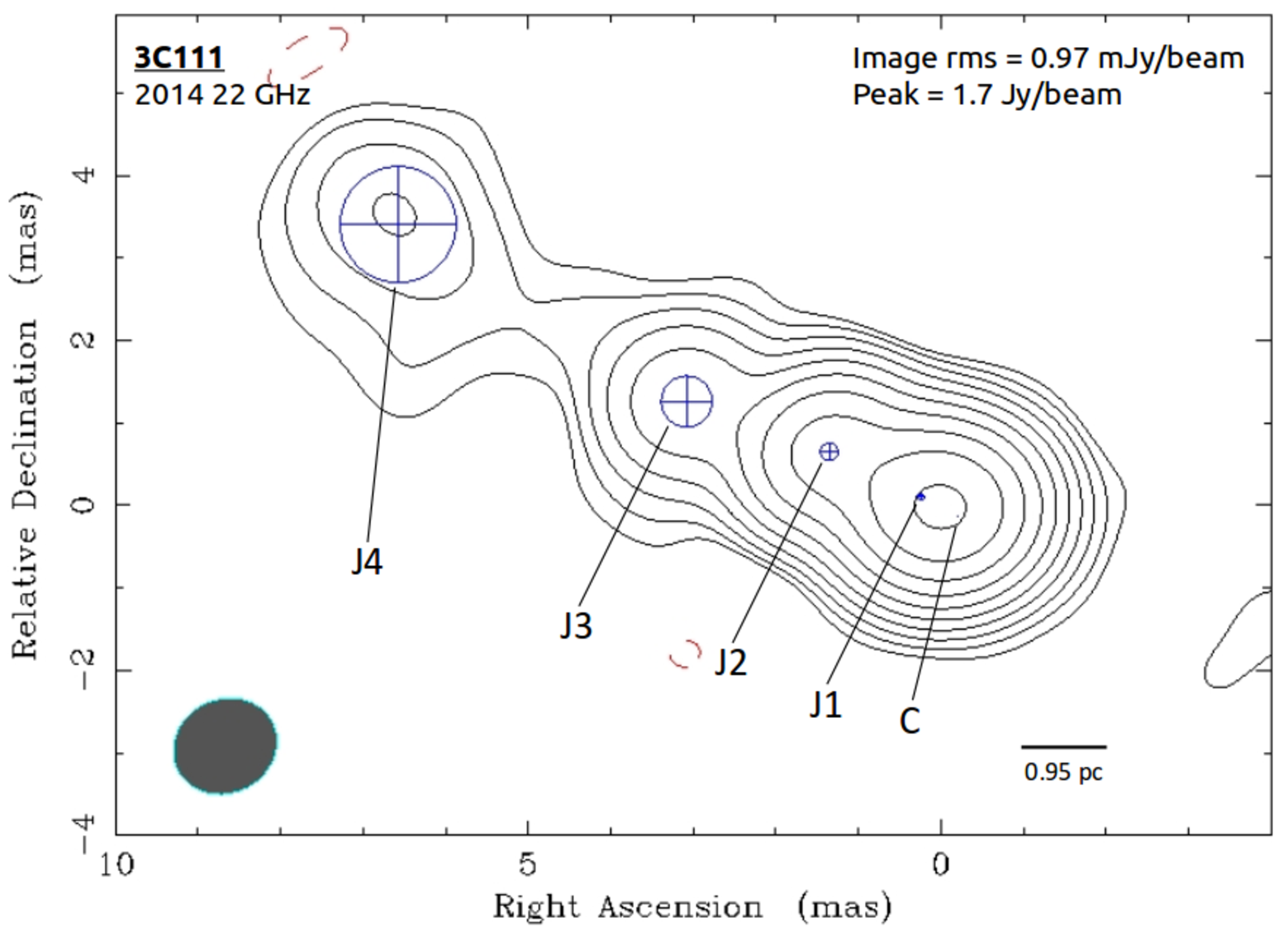}\hskip7mm 
\includegraphics[width=80mm]{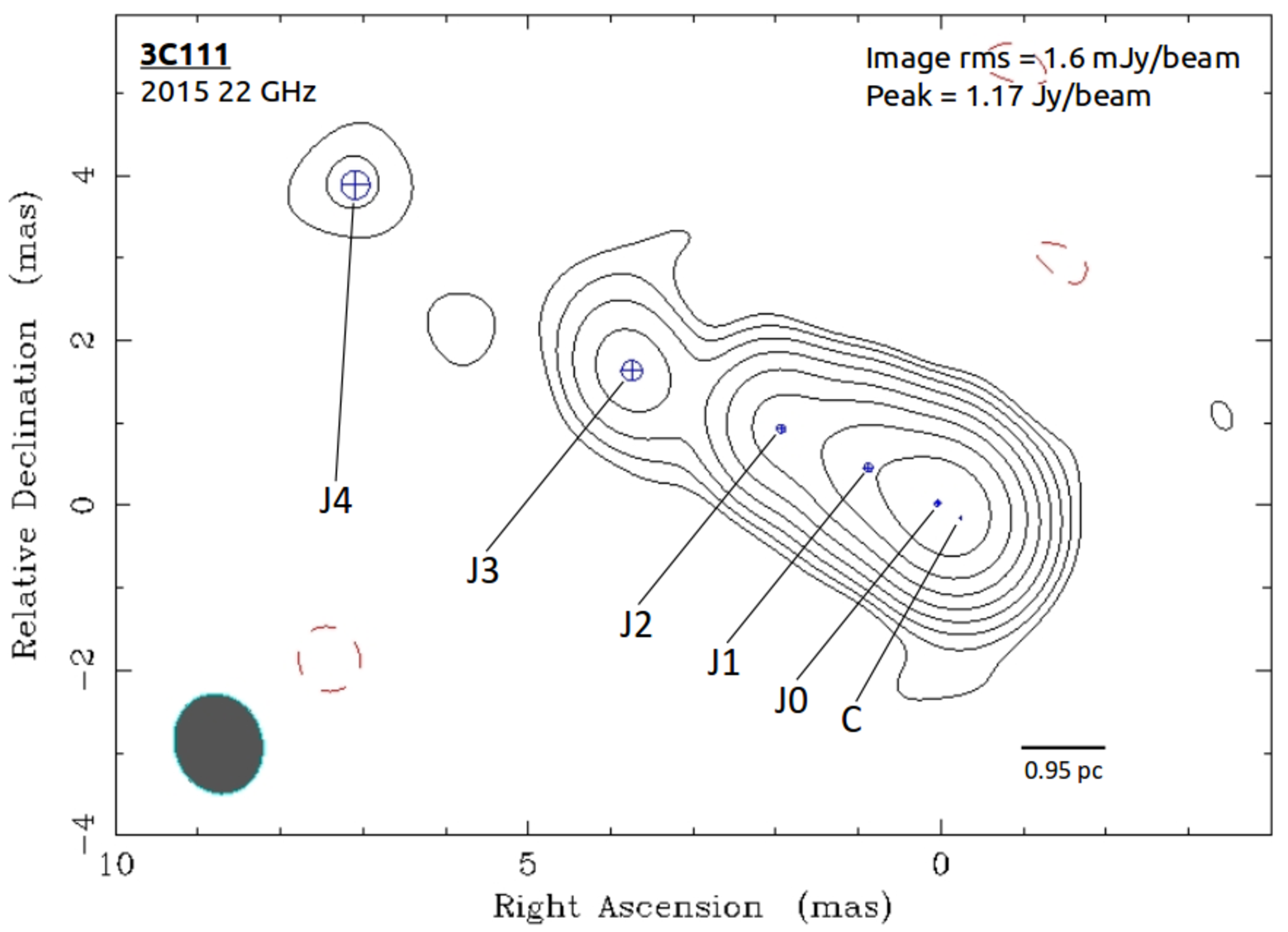} \\  
\includegraphics[width=80mm]{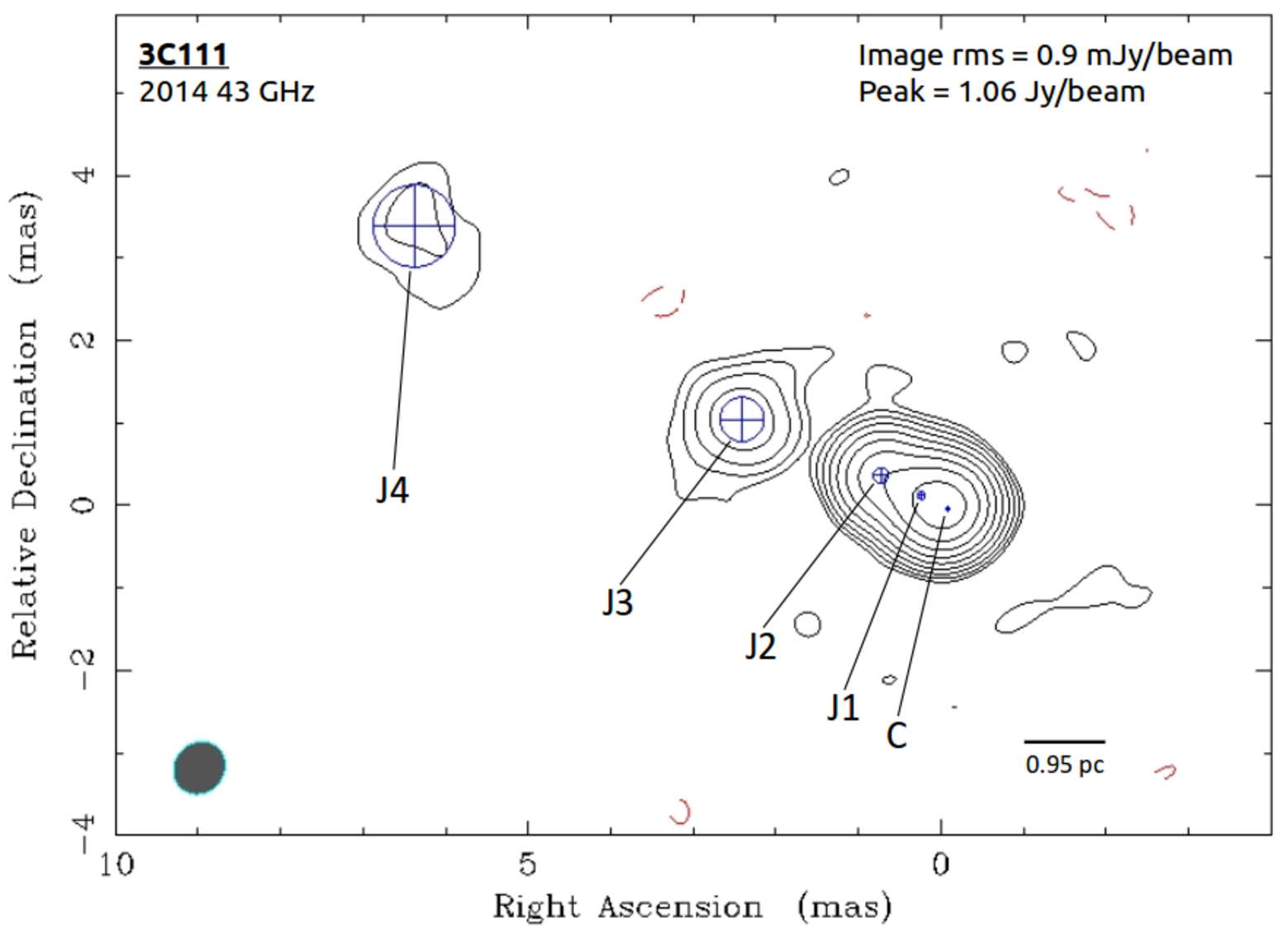}\hskip7mm 
\includegraphics[width=80mm]{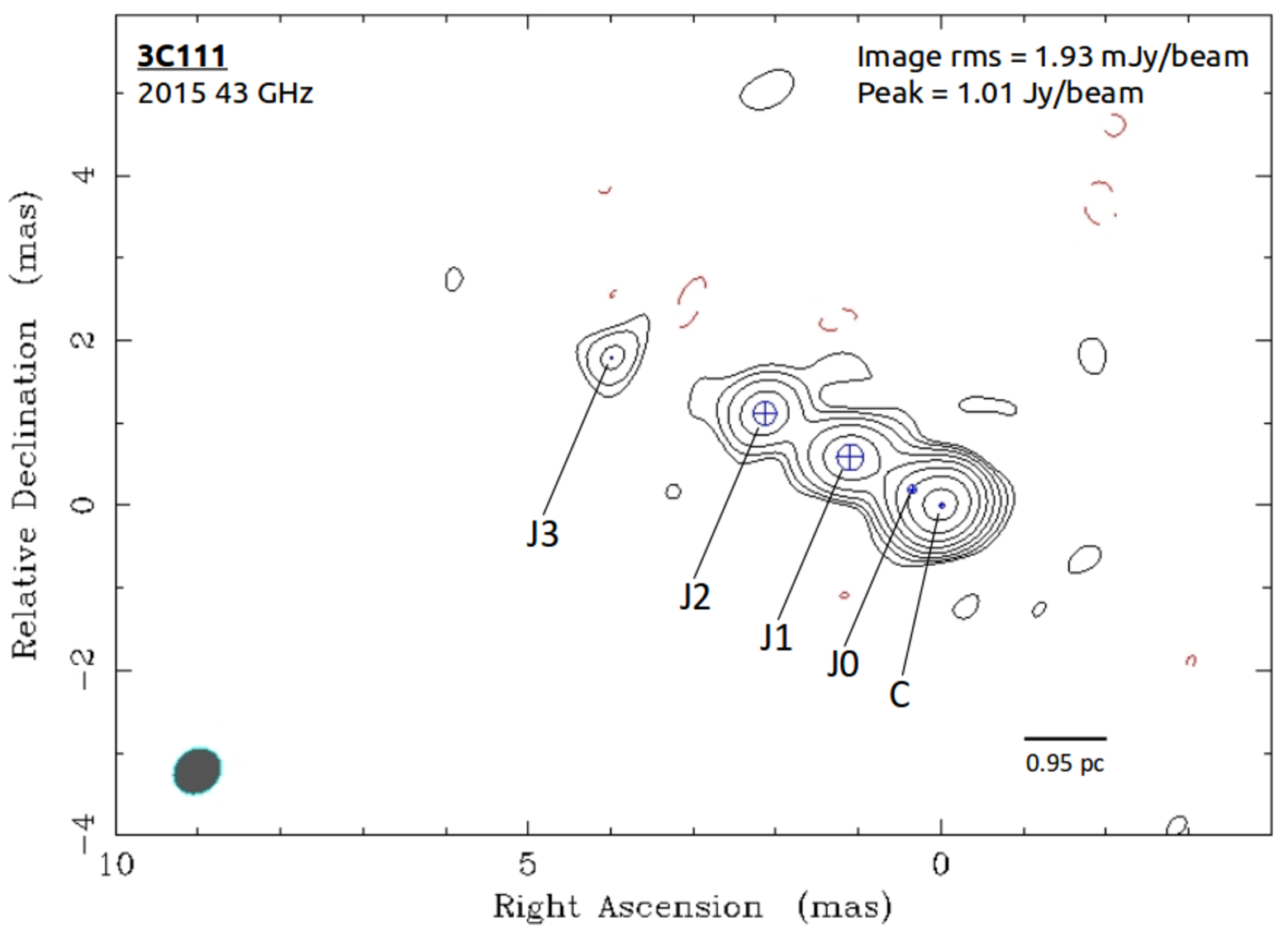}
\caption{Maps of 3C~111. \emph{Top left:} At 22 GHz in October 2014. \emph{Top right:} At 22 GHz in April 2015. \emph{Bottom left:} At 43 GHz in March 2014. \emph{Bottom right:} At 43 GHz in April 2015. Indicators are the same as in Figure~\ref{fig:fig1}.\label{fig:fig10}}
\end{figure*}

As for BL Lac, we obtained four images of 3C~111 in total (Figure~\ref{fig:fig10}). 3C~111 is a radio galaxy with a redshift of 0.491 \citep{Eracleous2004}. The corresponding angular scale is 0.95 pc/mas. A long jet, which extends up to $\sim$10 mas in north-east direction, can be found in each image. In the 43 GHz maps several knotty features are visible which are not distinguishable at 22 GHz. 5--6 model components can be found. The total flux of BL Lac was 2.64 Jy in 2014 and 1.93 in 2015 at 22 GHz, and 1.68 Jy and 1.5 Jy at 43 GHz.

\subsection{3C~345\label{sec:3c345}}

\begin{figure*}[t!]
\centering
\includegraphics[width=80mm]{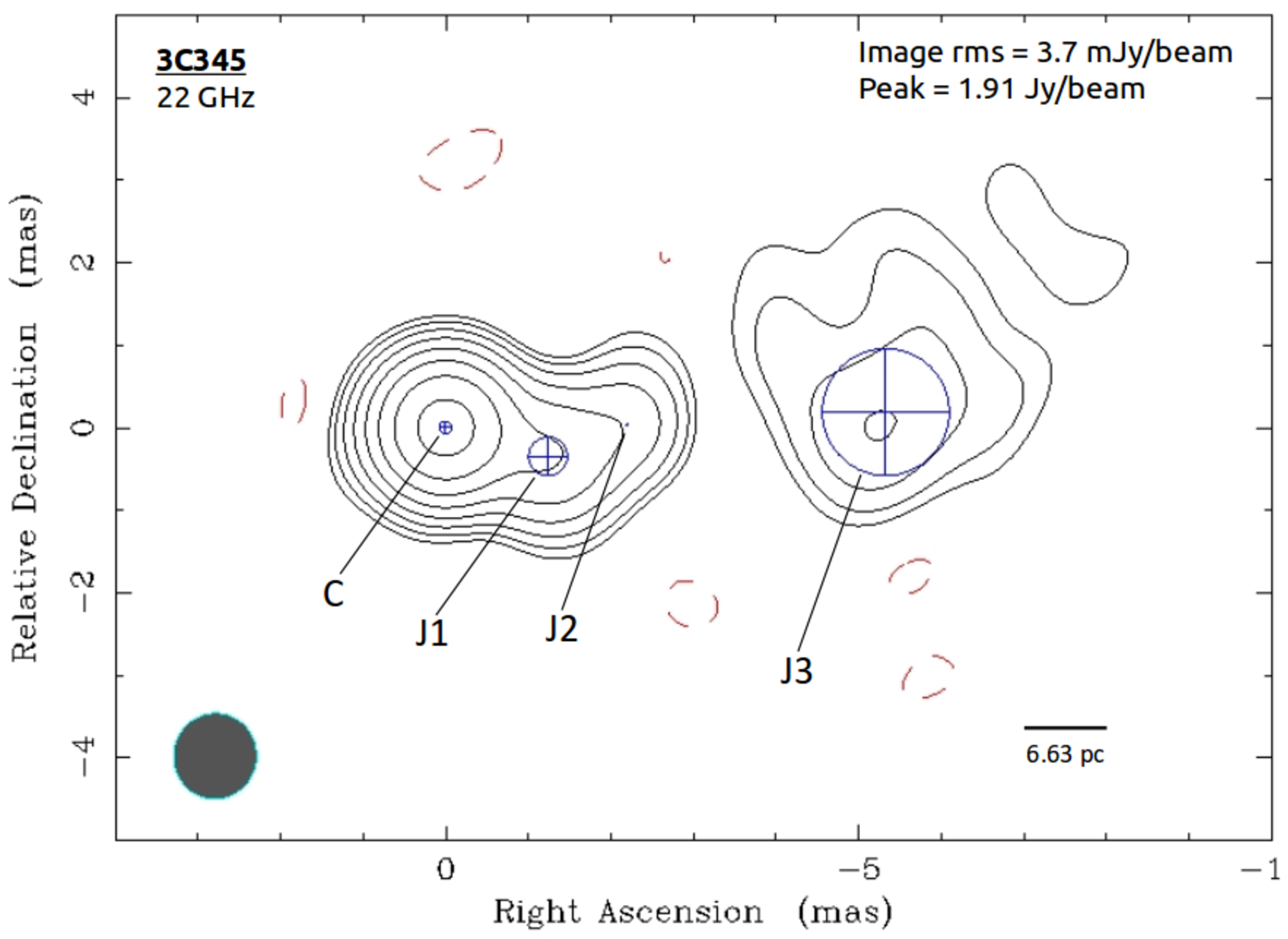}\hskip7mm
\includegraphics[width=80mm]{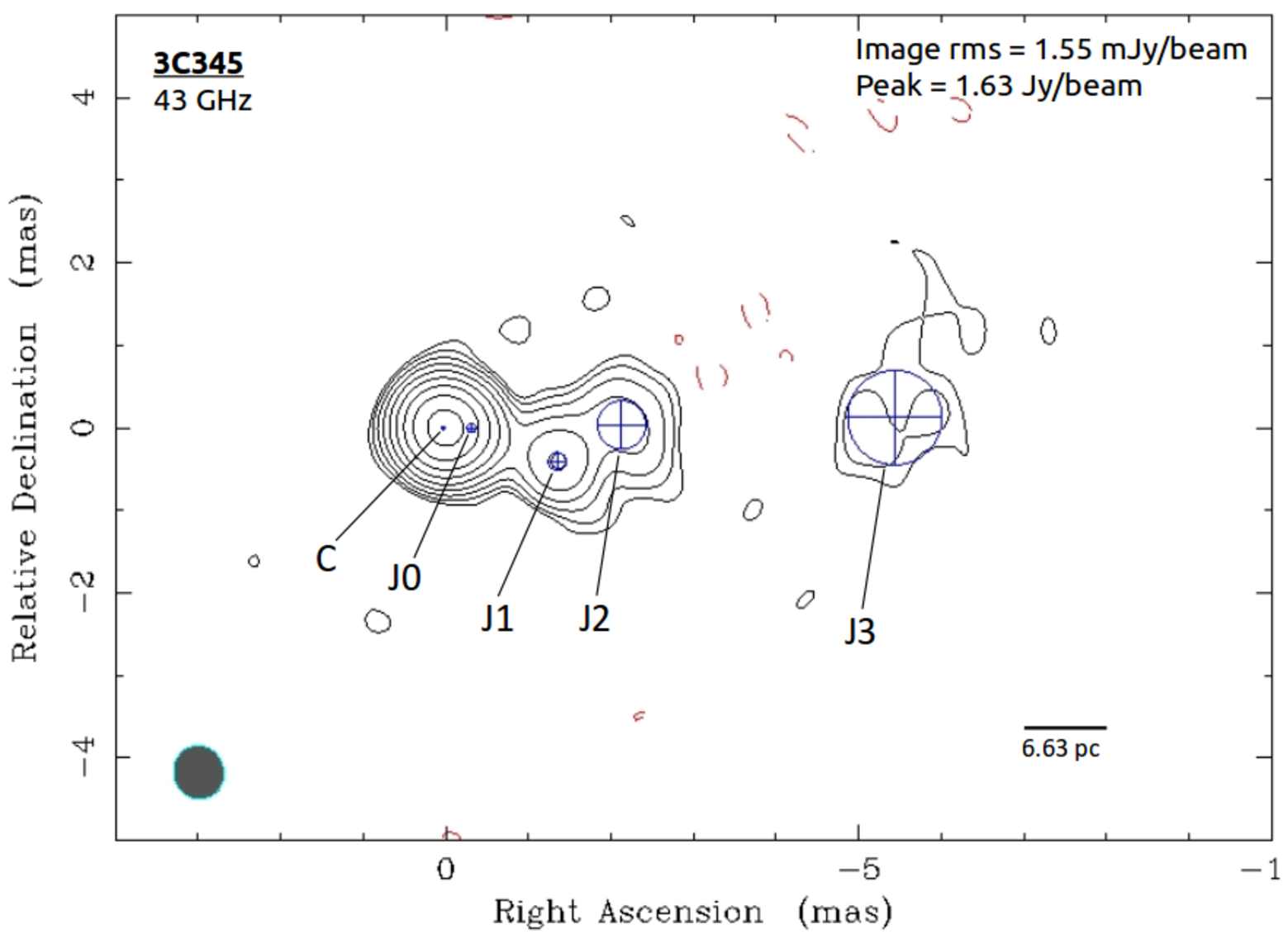}
\caption{Maps of 3C~345 obtained in March/April 2015. \emph{Left:} At 22 GHz. \emph{Right:} At 43 GHz. Indicators are the same as in Figure~\ref{fig:fig1}.\label{fig:fig14}}
\end{figure*}

3C~345 is a quasar with a redshift of 0.593 \citep{Marziani1996}. We included this source in our target list in 2015 as a replacement for 3C~84. Although 3C~345 has relatively high redshift compared to our other targets, it clearly shows its detailed jet structure in our KaVA maps. 3C~345 has been a ``popular'' object of study since the 1960s \citep[e.g.,][]{Goldsmith1965}; here, we focus on features like superluminal motion of the jet and evolution of the spectral index. 
Our results are shown in Figure~\ref{fig:fig14}. 3C~345 shows a complicated jet structure that extends out to $\sim$9 mas toward the west. The jet appears to bend at a distance of $\sim$1.5 mas from the core. The separate large jet component $\sim$5 mas west of the core shows a knotty structure. 3C~345 has a total flux of 2.76 Jy with a peak intensity of 1.91 Jy/beam at 22 GHz (2.07 Jy and 1.63 Jy/beam at 43 GHz). We were able to resolve four model components (including the core) at 22 GHz and found one more component at 43 GHz which seems to be a recently launched component that has not been resolved by the 22 GHz observations.

\section{Discussion\label{sec:disc}}

\subsection{Spectral Index\label{sec:spix}}

AGN jets are known to be emitters of continuous synchrotron emission which follows a powerlaw spectrum. The corresponding spectral index $\alpha$ is given by
\begin{equation}
\label{eq:spix}
S_\nu \propto \nu^\alpha
\end{equation}
where $S_\nu$ is the flux density observed at frequency $\nu$. The spectrum is approximately flat ($\alpha\gtrsim-0.5$) in optically thick emission regions and steep ($\alpha\lesssim-0.5$) in optically thin regions. For AGN in general, the core region is optically thick, while jets tend to become optically thinner as they expand on their way outward. 
Among our targets, BL Lac, 3C~111, and 3C~345 were observed simultaneously at both 22 and 43 GHz in 2015. We created a spectral index map for each source using the Python routine VIMAP which matches two radio maps obtained at different frequencies by using two-dimensional correlation \citep{Kim2014}. 

\begin{figure}[t!]
\centering
\includegraphics[trim=5mm 5mm 10mm 3mm, clip, width=80mm]{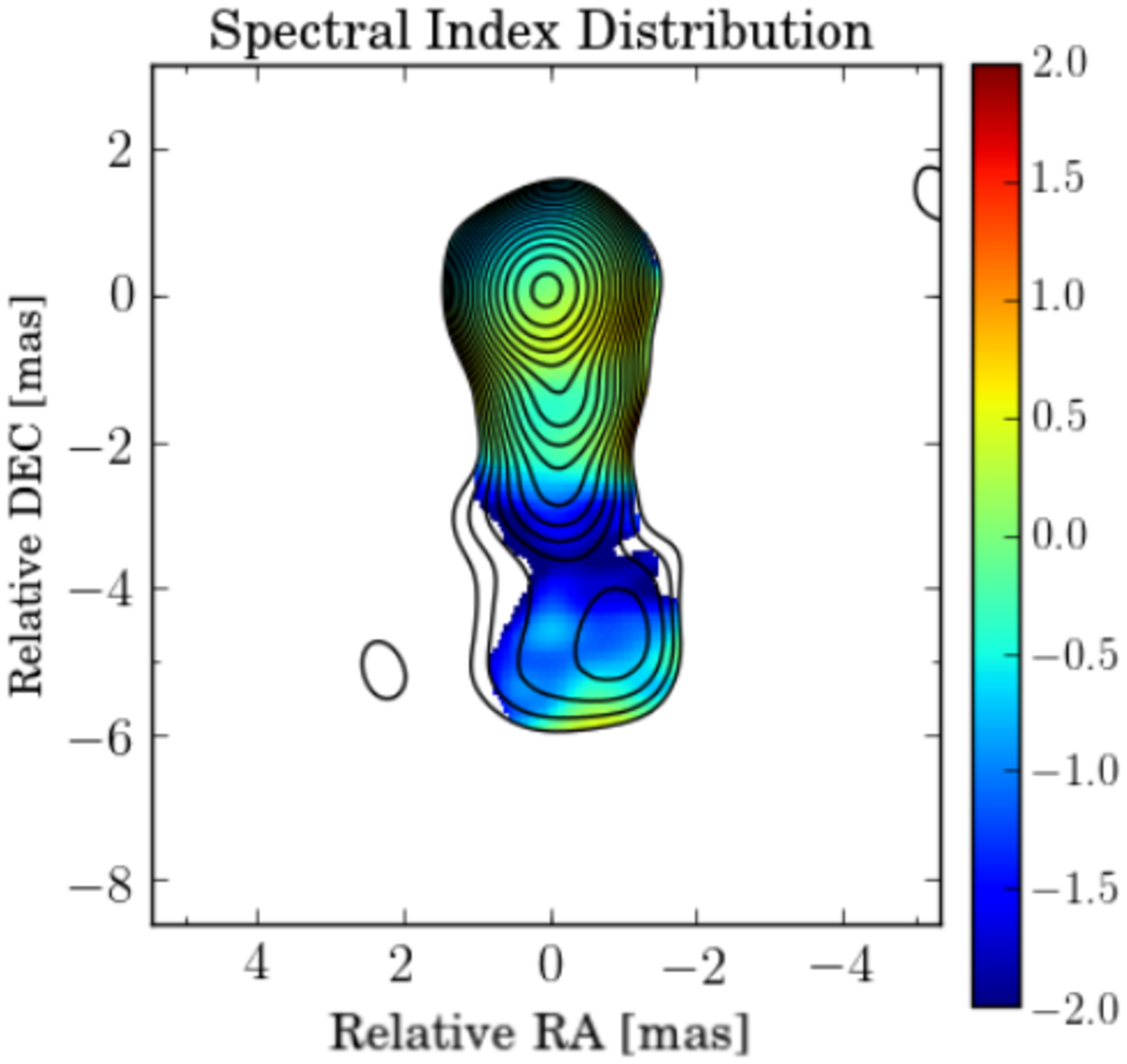}
\caption{Spectral index map of BL Lac. The color bar on the right indicates a spectral index $\alpha$ from $-2$ to $+2$.\label{fig:fig16}}
\end{figure}

\begin{figure}[t!]
\centering
\includegraphics[trim=1mm 5mm 10mm 3mm, clip, width=80mm]{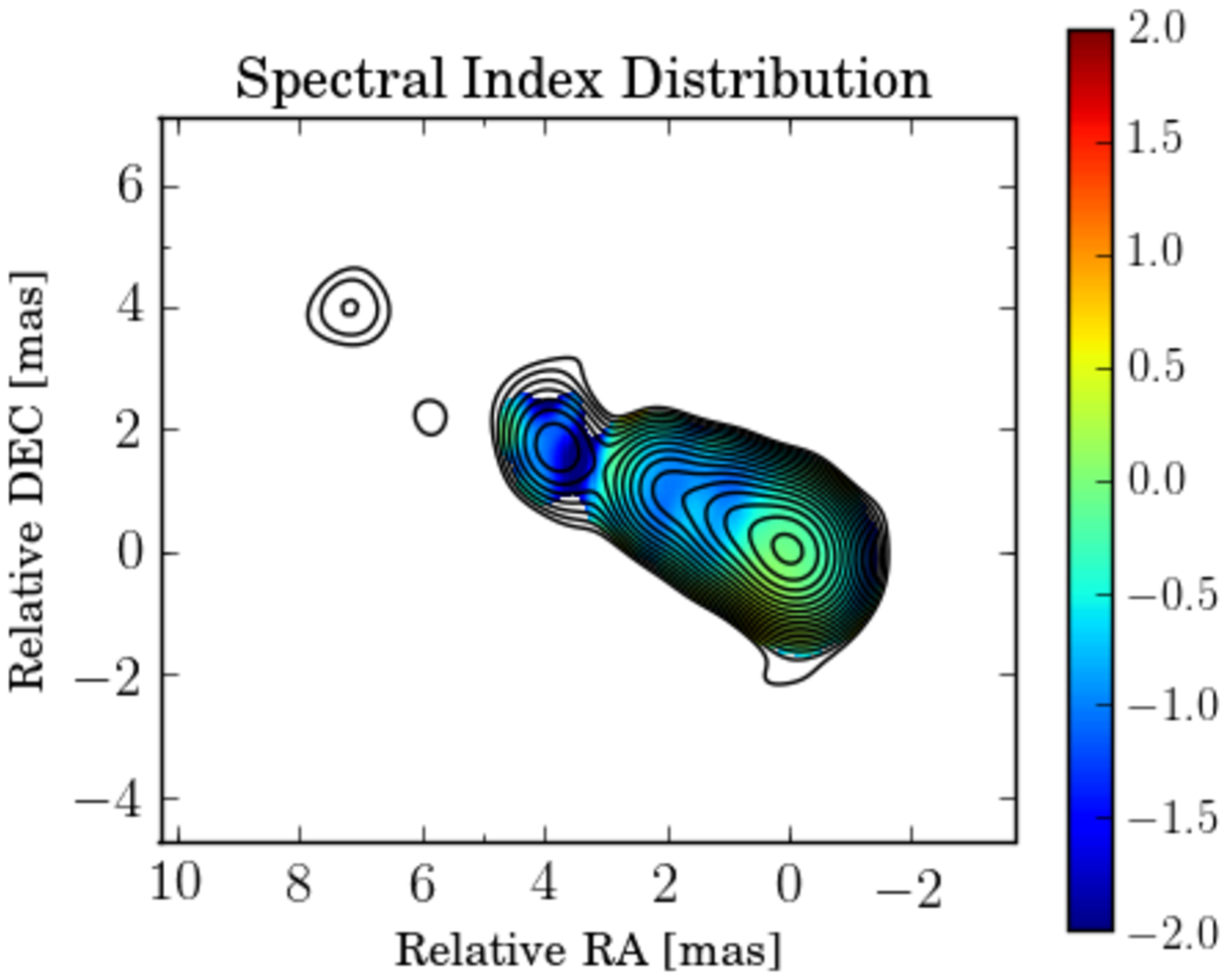}
\caption{Spectral index map of 3C~111. The color bar on the right indicates a spectral index $\alpha$ from $-2$ to $+2$. \label{fig:fig17}}
\end{figure}

\begin{figure}[t!]
\centering
\includegraphics[trim=1mm 5mm 10mm 3mm, clip, width=80mm]{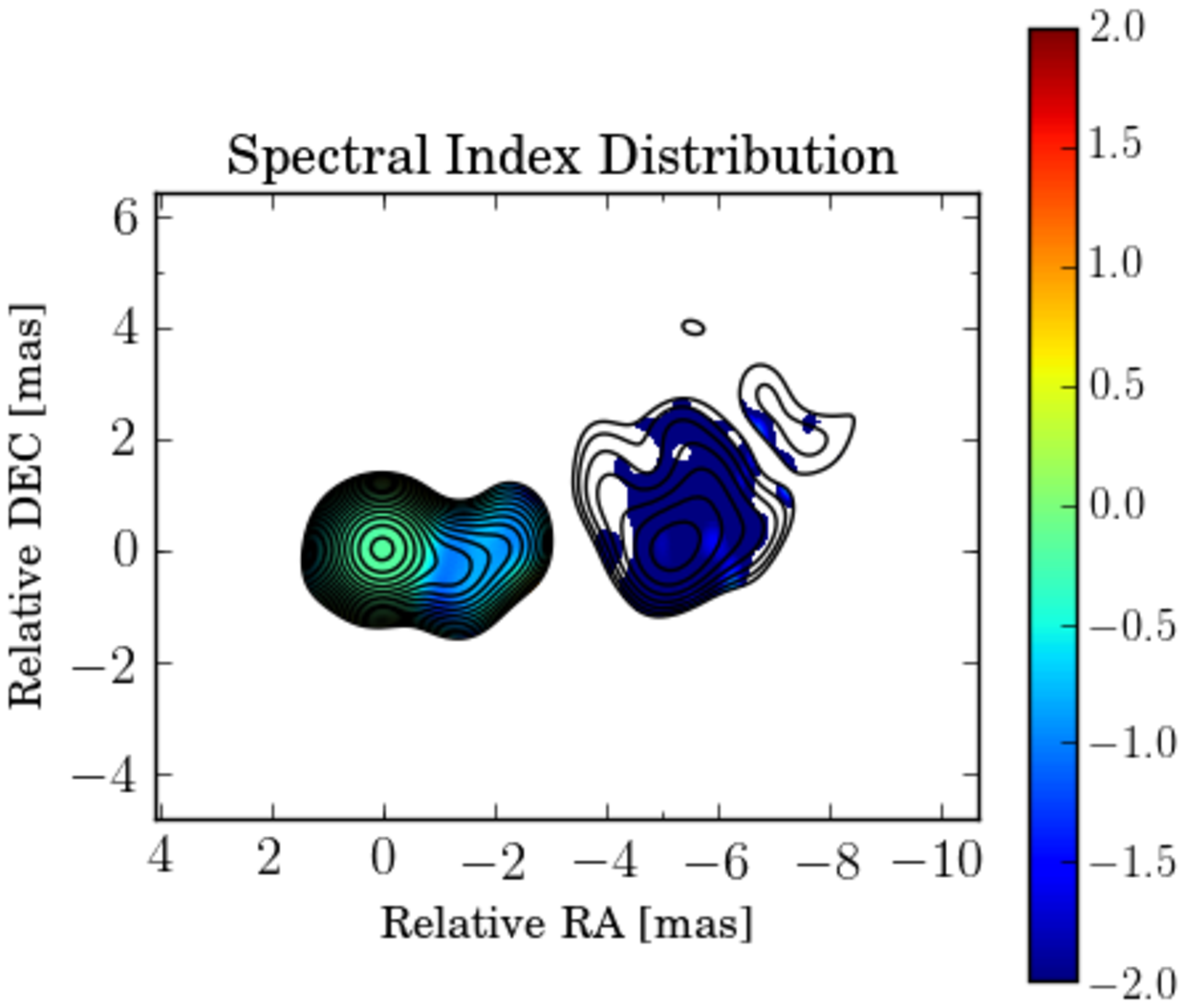}
\caption{Spectral index map of 3C~345. The color bar on the right indicates a spectral index $\alpha$ from $-2$ to $+2$. \label{fig:fig18}}
\end{figure}

Figure \ref{fig:fig16}, \ref{fig:fig17}, and \ref{fig:fig18} are the spectral index maps for all three sources using the 22 and 43 GHz maps. We assumed that the core-shift effect \citep{Porcas2009} is negligible at our observing frequencies. In general, as expected, the maps show flat spectral index values in the core regions and increasingly steep indices along the jets. This is especially obvious in case of 3C~345, with values as small as $\alpha\approx-2$ in the outermost parts of the jet. The steep spectral index in jet of 3C~345 is also consistent with previous studies \citep[e.g.,][]{Ros2000, Lobanov1999, Rantakyroe1995}. 
Coincidentally, BL Lac, 3C~111, and 3C~345 correspond to three different AGN types (a BL Lac object, a radio galaxy, and a quasar, respectively). Initially, all sources show flat cores. BL Lac shows $\alpha\approx0$ also along the jet down to $\sim$3 mas from the core, 3C~111 has a moderately steep ($\alpha \approx -1$) spectral index throughout the jet, and 3C~345 shows a very steep ($\alpha \approx -2$) outermost jet component. This sequence -- flat (BL Lac object) $\to$ moderate steep (radio galaxy) $\to$ steep (quasar) -- could be caused by different viewing angles.

\subsection{Kinematics of Jet Components\label{sec:kinematics}}

In the cases of BL Lac and 3C~111, we were able to trace the motion of their jets between 2014 and 2015.

\subsubsection{BL Lac}

\begin{figure}[t!]
\centering
\includegraphics[width=80mm]{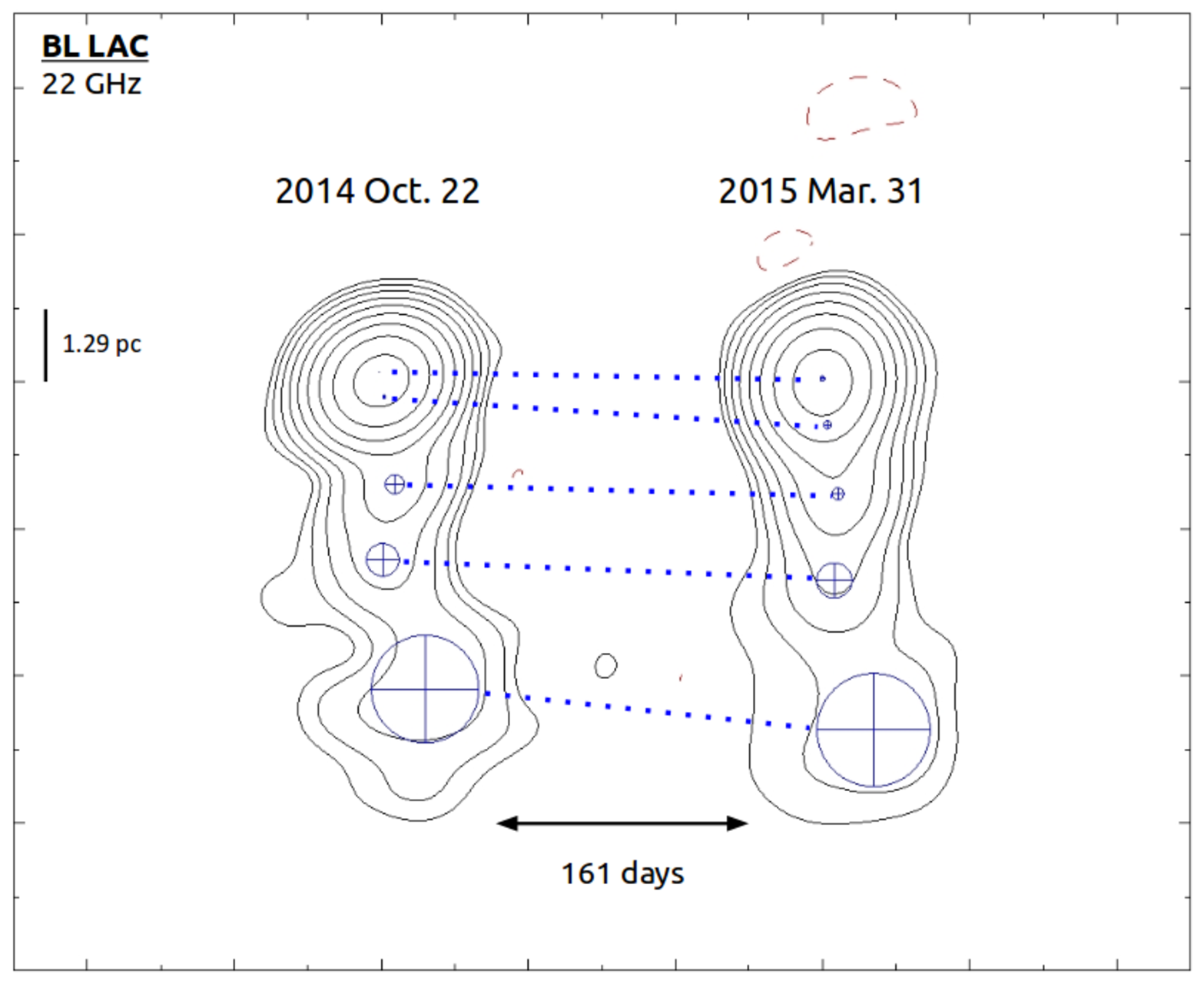}
\caption{Evolution of BL Lac between 2014 and 2015 at 22 GHz. Five model components can be matched (blue dashed lines).\label{fig:fig19}}
\end{figure}

We matched core and jet components in the BL Lac 22 GHz data (Figure~\ref{fig:fig19}) and found the jet components propagating toward the south. The time between both observations was 161 days. Components J1--J4 moved between 0.067 and 0.62 parsec, which corresponds to apparent speed from $0.5c$ to $4.6c$ (Table~\ref{tab:table11}).

The relation between the apparent speed $\beta_{\rm app}$ and the jet orientation $\theta$ can be expressed by 
\begin{equation}
\label{eq:beta}
\beta_{\rm app} = \frac{\beta \sin \theta}{1 - \beta \cos \theta}
\end{equation}
where $\beta$ is the intrinsic velocity in units of the speed of light and $\theta$ is the angle between the line of sight and the jet \citep[e.g.,][]{Kellermann2007}. We show the apparent speed as function of viewing angle in Figure~\ref{fig:fig21} for several values of $\beta$ (0.5, 0.9, 0.95, 0.98, 0.99, 0.995, 0.998 and 0.999). As one can read off the diagram in a straightforward manner, an apparent speed $\beta_{\rm app}\approx 5c$ (dashed line in the diagram) implies a lower limit for the intrinsic speed of $\sim0.98c$ (lime line) and an upper limit of the viewing angle of $\sim20\deg$. We note the ambiguity in viewing angle which could have very small values of a few degrees. However, we consider such small angles unlikely for BL Lac (and also for 3C~111 -- cf. Section \ref{sec:3c111speed}) because we would expect the jet to be optically thick along its entire length in this case, in contradiction to the distribution shown by the spectral index map.

In the 43 GHz data, we were not able to cross-identify individual components due to poor quality of the 2014 data. Furthermore, substantial changes in both morphology and physical properties (like total flux: $\sim$40\% decreased; peak intensity: $\sim$50\% decreased) occurred within the 350 days passed between the two 43 GHz datasets, further complicating the comparison.

\subsubsection{3C~111 \label{sec:3c111speed}}

\begin{figure*}[t!]
\centering
\includegraphics[width=67mm]{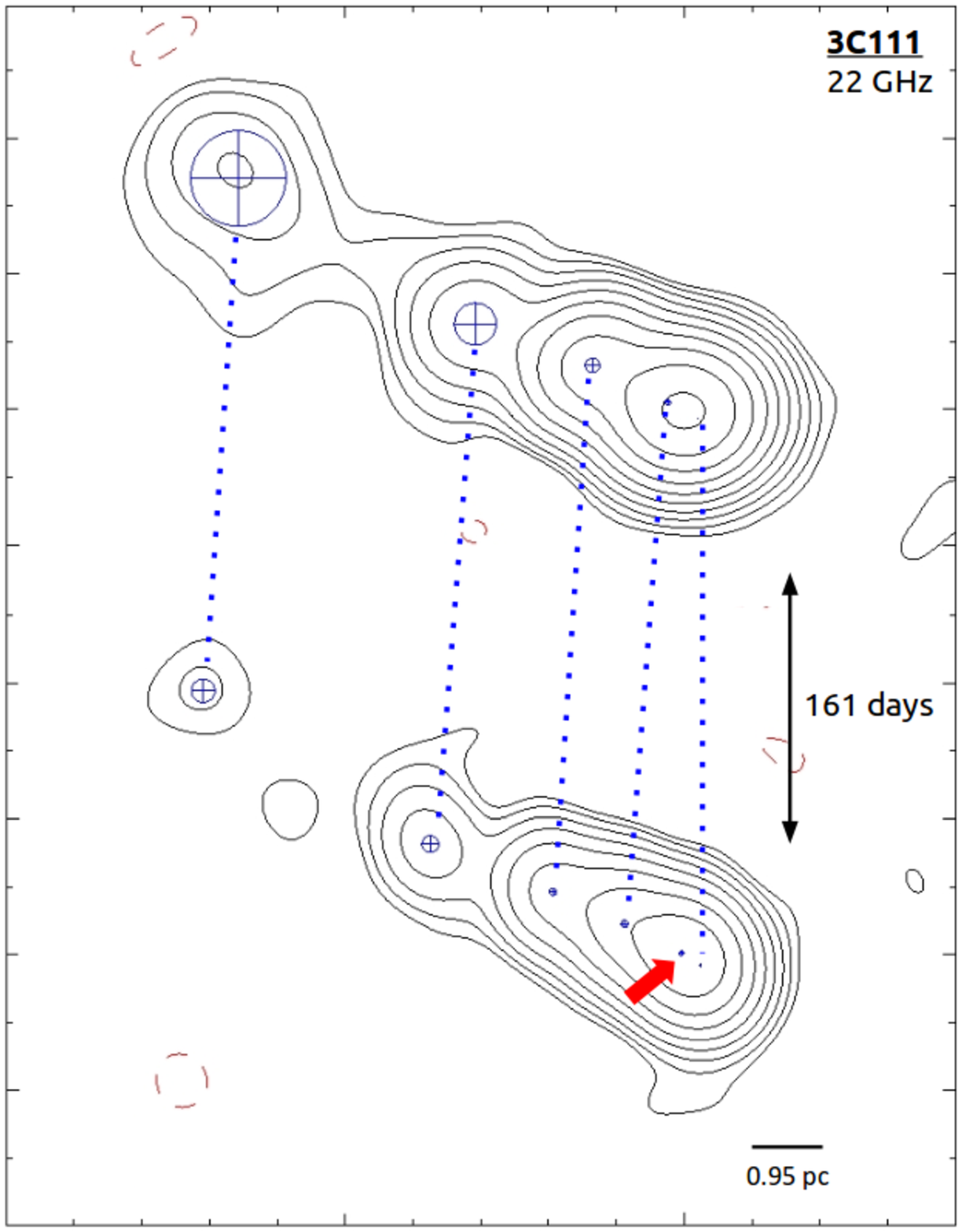}\hskip7mm
\includegraphics[width=67mm]{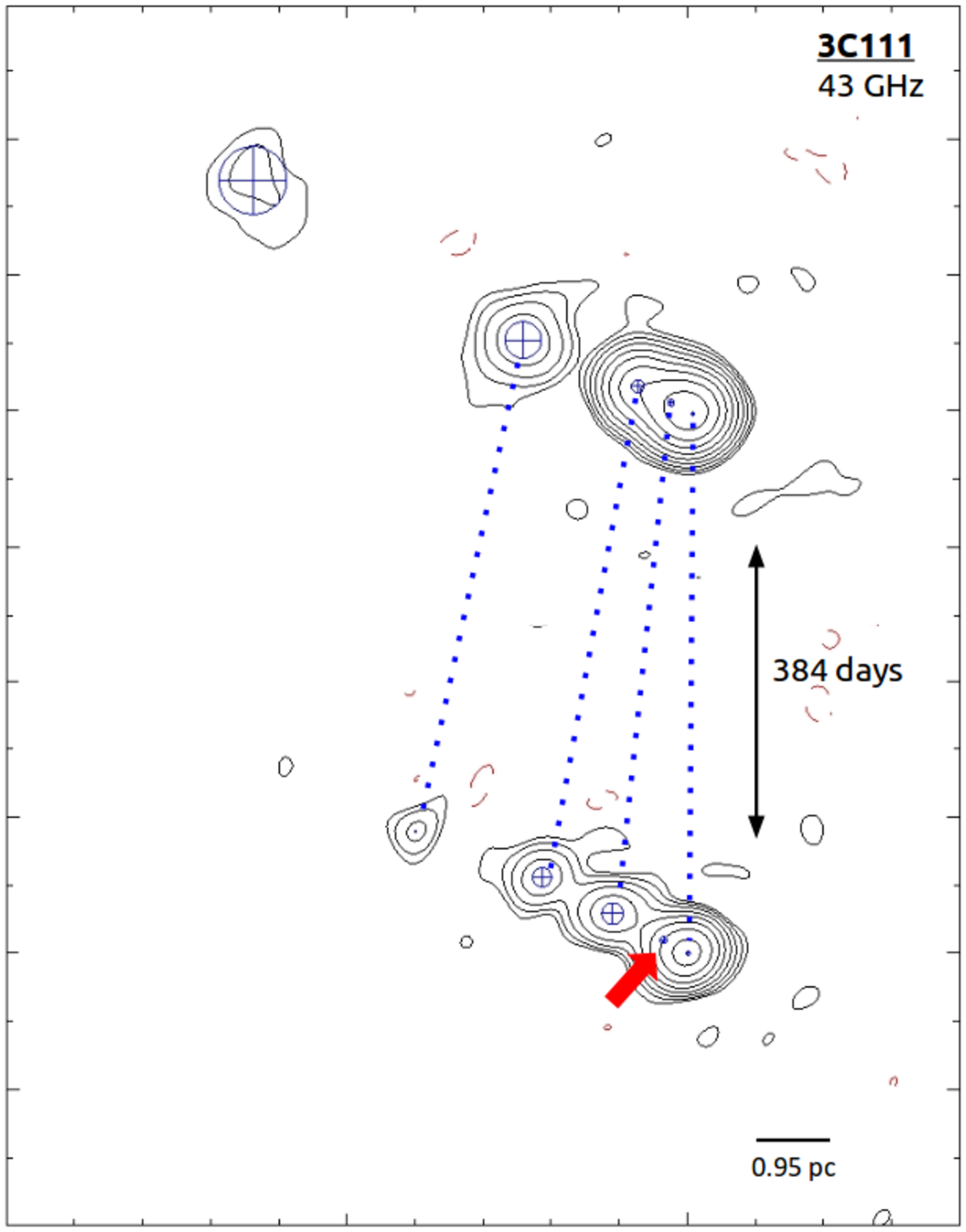} \\[3mm]  
\includegraphics[width=67mm]{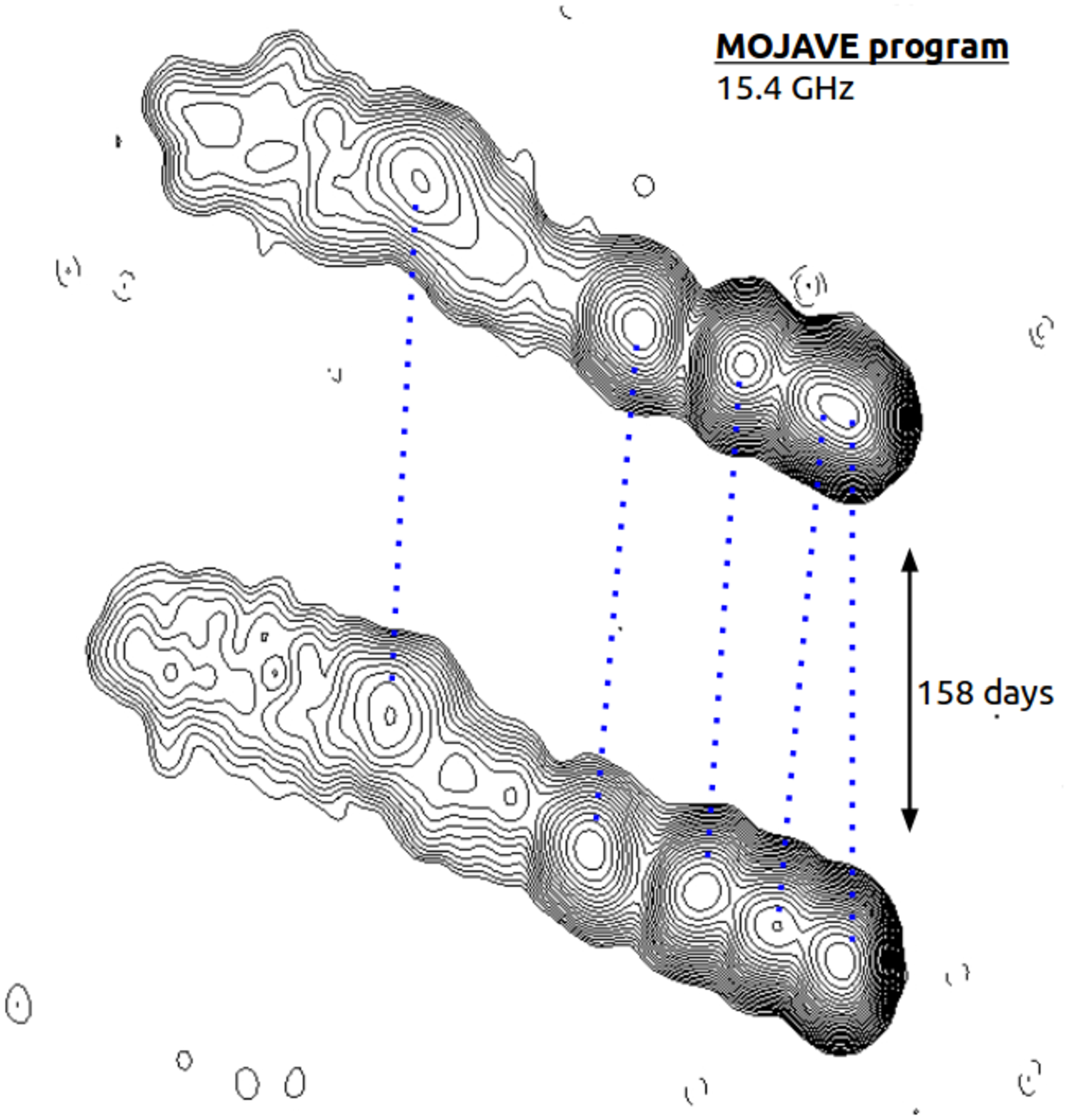}\hskip7mm
\includegraphics[width=67mm]{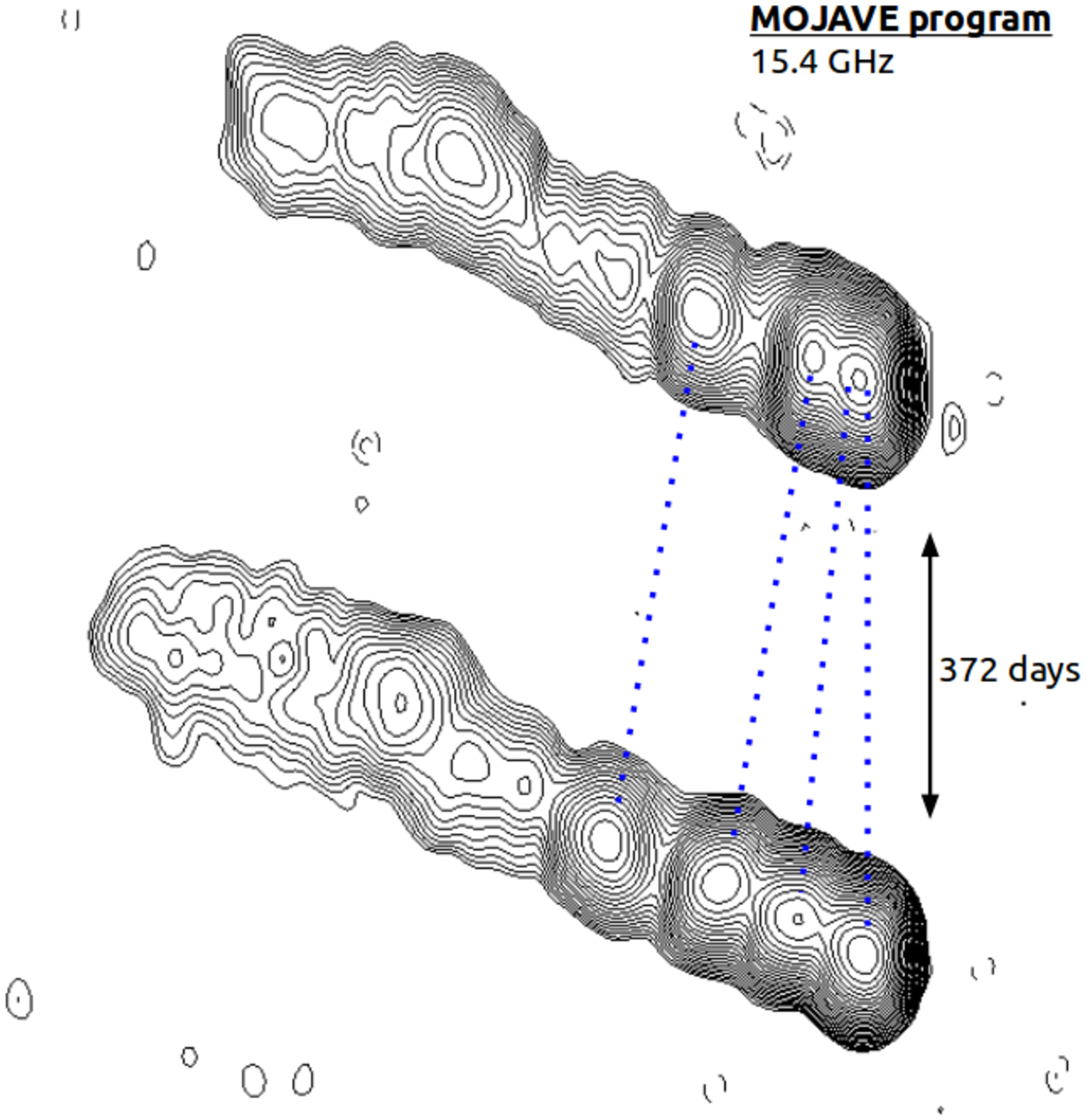}
\caption{Time-resolved VLBI maps of 3C~111 obtained by us with KaVA (top panels) and by the MOJAVE program with the VLBA (bottom panels).
\emph{Top left:} Comparison of 22-GHz KaVA maps from 2014 and 2015. We assumed that a new jet component was launched in 2015 (red arrow, J0 in Figure~\ref{fig:fig10} top right), and matched five model components visible in both epochs (blue dashed lines).
\emph{Top right:} Comparison of 43-GHz KaVA maps from 2014 and 2015. We assumed that a new jet component ws in 2015 (red arrow, J0 in Figure~\ref{fig:fig10} bottom right), and matched four model components visible in both epochs (blue dashed lines).
\emph{Bottom left:} MOJAVE images obtained at 15.4 GHz, observed on 12 December 2014 (top) and 18 May 2015 (bottom). We re-identified and matched the same jet components as in the 22-GHz KaVA maps.
\emph{Bottom right:} MOJAVE images obtained at 15.4 GHz, observed on 12 May 2014 (top) and 18 May 2015 (bottom). We re-identified and matched the same jet components as in the 43 GHz KaVA maps. The results by KaVA and MOJAVE are in good agreement.
\label{fig:fig20}}
\end{figure*}

\begin{figure}[t!]
\centering
\includegraphics[width=80mm]{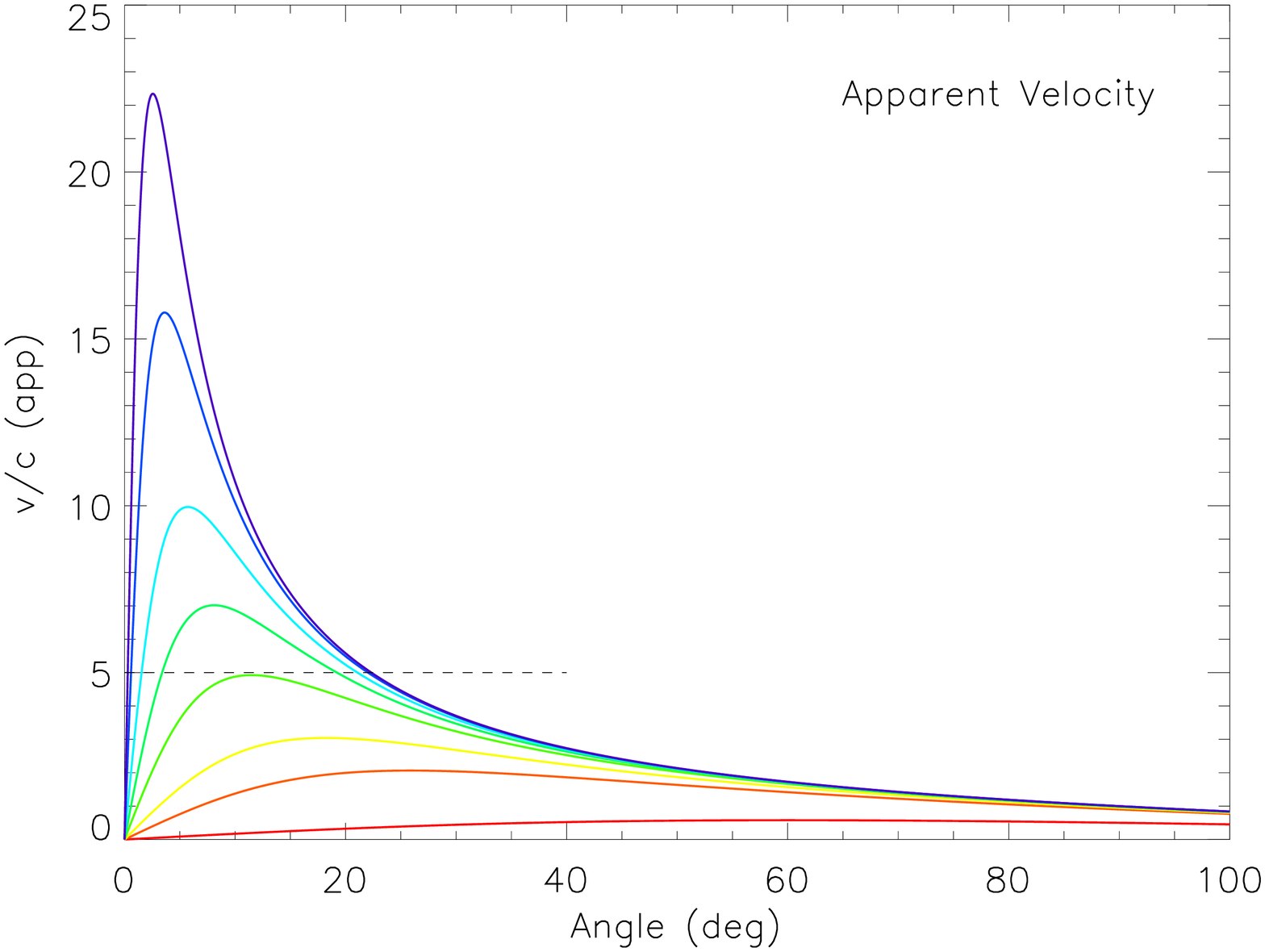}
\caption{Apparent speed as a function of jet orientation for various values of $\beta$. From the top, $\beta$=0.999 (purple), $\beta$=0.998 (blue), $\beta$=0.995 (cyan), $\beta$=0.99 (green), $\beta$=0.98 (lime), $\beta$=0.95 (yellow), $\beta$=0.9 (orange) and $\beta$=0.5 (red). The black dashed line,  which touches the peak of the lime line ($\beta$=0.98), indicates an apparent speed of $5c$.\label{fig:fig21}}
\end{figure}

Figure~\ref{fig:fig20} shows the comparison between observations of 3C~111 obtained in 2014 and 2015. At 22 GHz, the time gap between the two observations is 161 days. We assumed that the first model component (marked by a red arrow) in the 2015 map, which is located just next to the core, is a newly launched jet component. We matched the remaining four components and calculated the distance over which they moved. Components J0--J4 moved between 0.31 and 0.76 parsec, corresponding to a (average) apparent speed of around $4.6c$ (Table~\ref{tab:table11}).

We applied the same approach to our 43 GHz data. We identified component J0 in the 2015 map as a newly launched jet component and matched the components J1--J3. J4 was not detected in 2015. At 43 GHz, the time span between the maps is 384 days. Else than for the 22 GHz data, we find travel distances ranging from 0.93 pc to 1.66 pc; the further the component is located from the core, the larger distance it moved. Apparent speeds range from $2.9c$ to $5.2c$. Our results

\begin{table}[t!]
\caption{Apparent speed of jet components of BL Lac and 3C~111\label{tab:table11}}
\centering
\begin{tabular}{lrr}
\toprule
ID & Distance moved & $\beta_{\rm app}$  \\
   &   (pc)                 & ($c$) \\
\midrule
{\sc BL Lac 22 GHz:} & &  \\
J1 & 0.385	& 2.85 \\
J2 & 0.067	& 0.49 \\
J3 & 0.301	& 2.23 \\
J4 & 0.620	& 4.59 \\ \addlinespace
{\sc 3C~111 22 GHz:} & &  \\
J0 & 0.312 & 2.31 \\
J1 & 0.728 & 5.38 \\
J2 & 0.647 & 4.78 \\
J3 & 0.757 & 5.60 \\
J4 & 0.705 & 5.21 \\ \addlinespace
{\sc 3C~111 43 GHz:} & &  \\
J0 & 0.387 & 1.20 \\
J1 & 0.845 & 2.62 \\
J2 & 1.431 & 4.44 \\
J3 & 1.578 & 4.89 \\
\bottomrule
\end{tabular}
\end{table}

In order to confirm the motion of the jets and to check our analysis, we repeated our analysis with 15.4 GHz VLBA data from the MOJAVE program. Although observation frequencies and epochs are different, we were able to make a reasonable comparison to our data. We picked MOJAVE maps obtained on 2014 May 12, 2014 December 12, and 2015 May 18, which is about one month delayed compared to our observations. The time spans covered are 158 and 372 days, respectively. The propoer motions of jet components in the MOJAVE maps (Figure~\ref{fig:fig20}, bottom panels) are consistent with our results based on KaVA data.

\subsection{Jet Expansion \label{sec:expand}}

\begin{figure}[t!]
\centering
\includegraphics[trim=8mm 2mm 12mm 8mm, clip, width=80mm]{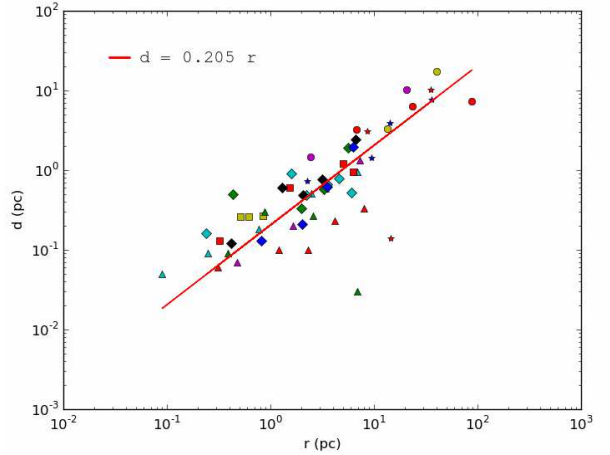}
\caption{Jet component size $d$ vs. distance from core $r$, in units of parsec, for all available data sets. Markers indicate 1055+018 (yellow circle), 0133+476 (magenta circle), 1642+690 (red circle), 3C~120 (red square), 3C~84 (yellow square), 3C~111 (triangle: magenta -- 2014 22 GHz, cyan -- 2014 43 GHz, red -- 2015 22 GHz, green -- 2015 43 GHz), BL Lac (diamond: green -- 2014 22 GHz, cyan -- 2014 43 GHz, blue -- 2015 22 GHz, black -- 2015 43 GHz) and 3C~345 (star: red -- 22 GHz, blue -- 43 GHz). The data follow a global relation $d\approx0.2r$.\label{fig:dvsr}}
\end{figure}

As noted several times throughout Section~\ref{sec:result}, our maps indicate a trend toward more extended jet components at larger core distances, as expected for expanding jets. When combining the data for all sources at all frequencies and all epochs in a common diagram, one finds the result illustrated in Figure~\ref{fig:dvsr}. Indeed, one finds larger component sizes $d$ for larger core distances $r$. Remarkably, all data approximately follow the same relation $d\approx0.2r$, albeit within large scatter. This is in good agreement with global jet expansion; however, we note that this relation comes with substantial systematic uncertainties because we average over sources with different viewing angles and jet kinematics. At the very least, we demonstrated that already few epochs of KaVA data are able to identify universal properties of AGN jets.


\begin{table*}[p!]
\caption{Parameters for model components\label{tab:table12}}
\centering
\begin{tabular}{lcccccccccc}
\toprule
Source & $S_{Tot}$ & $\sigma_{Tot}$ & $S_{Peak}$ & $\sigma_{Peak}$ & $r$ & $\sigma_{r}$ & $\theta$ & $\sigma_{\theta}$ & $d$ & $\sigma_{d}$  \\
ID    & (Jy) & (mJy) & (Jy/beam) & (mJy/beam) & (pc) & (pc) & ($\deg$) & ($\deg$) & (pc) & (pc) \\
\midrule
{\sc 1055+018} & & & & & & & & & & \\
22 GHz: & & & & & & & & & & \\
C	&	3.64	&	176.04	&	3.62	&	87.75	&	---	&	---	&	---	&	---	&	0.17	&	0.032	\\
J1	&	0.08	&	14.3	&	0.07	&	6.93	&	13.53	&	0.16	&     $-$65.91	&	0.012	&	0.43	&	0.31	\\
J2	&	0.11	&	27.22	&	0.04	&	7.43	&	40.41	&	1.62	&     $-$57.09	&	0.04	&	2.25	&	3.25	\\
\midrule 
{\sc 0133+476} & & & & & & & & & & \\
22 GHz: & & & & & & & & & & \\			
C	&	1.35	&	103.88	&	1.35	&	52	&	---	&	---	&	---	&	---	&	0.03	&	0.009	\\
J1	&	0.32	&	44.57	&	0.32	&	22.04	&	2.45	&	0.05	&      $-$76.79	&	0.021	&	0.19	&	0.1	\\
J2	&	0.046	&	13.7	&	0.023	&	4.56	&	20.63	&	1.02	&      $-$35.46	&	0.049	&	1.33	&	2.04	\\
\midrule 
{\sc 1642+690} & & & & & & & & & & \\	
22 GHz: &&&&&&&&&& \\  
C	&	0.51	&	31.99	&	0.51	&	15.99	&	---	&	---	&	---	&	---	&	0.09	&	0.021	\\
J1	&	0.35	&	40.35	&	0.31	&	19.1	&	6.81	&	0.1	&     $-$156.63	&	0.015	&	0.44	&	0.2	\\
J2	&	0.004	&	3.88	&	0.004	&	1.89	&	23.58	&	1.52	&     $-$167.9	&	0.065	&	0.85	&	3.05	\\
J3	&	0.006	&	4.68	&	0.004	&	2.03	&	88.36	&	1.67	&     $-$166.61	&	0.019	&	0.99	&	3.34	\\
\midrule 
{\sc 3C~120} & & & & & & & & & & \\ 
22 GHz: &&&&&&&&&& \\ 
C	&	1.31	&	78.58	&	1.31	&	39.34	&	---	&	---	&	---	&	---	&	0.04	&	0.001	\\
J1	&	0.22	&	17.37	&	0.21	&	8.62	&	0.32	&	0.003	&     $-$106.39	&	0.008	&	0.2	&	0.005	\\
J2	&	0.08	&	19.85	&	0.05	&	8.08	&	1.54	&	0.05	&     $-$112.45	&	0.029	&	0.94	&	0.09	\\
J3	&	0.021	&	12.8	&	0.009	&	3.87	&	5.04	&	0.25	&     $-$113.98	&	0.05	&	1.88	&	0.51	\\
J4	&	0.019	&	10.48	&	0.009	&	3.39	&	6.29	&	0.18	&     $-$111.43	&	0.029	&	1.46	&	0.36	\\
\midrule 
{\sc 3C~84} & & & & & & & & & & \\
22 GHz: &&&&&&&&&& \\  
C	&	4.1	&	460.55	&	3.71	&	218.88	&	---	&	---	&	---	&	---	&	0.43	&	0.009	\\
J1	&	2.55	&	442.3	&	1.89	&	188.42	&	0.51	&	0.013	&	170.09	&	0.025	&	0.74	&	0.026	\\
J2	&	1.64	&	247.01	&	1.22	&	105.25	&	0.62	&	0.011	&    $-$153.39	&	0.018	&	0.75	&	0.023	\\
J3	&	13.39	&	1301.13	&	9.71	&	546.98	&	0.84	&	0.008	&	177.33	&	0.009	&	0.78	&	0.015	\\
\midrule 
{\sc BL Lac 2014} & & & & & & & & & & \\
22 GHz: &&&&&&&&&& \\ 
C	&	1.75	&	84.72	&	1.75	&	42.38	&	---	&	---	&	---	&	---	&	0.01	&	0.0002	\\
J1	&	1.02	&	108.17	&	1.02	&	54.01	&	0.43	&	0.001	&    $-$170.06	&	0.003	&	0.04	&	0.003	\\
J2	&	0.17	&	29.22	&	0.16	&	14.18	&	1.97	&	0.015	&    $-$172.36	&	0.008	&	0.26	&	0.03	\\
J3	&	0.09	&	21.77	&	0.074	&	9.87	&	3.28	&	0.038	&    $-$179.1	&	0.012	&	0.44	&	0.076	\\
J4	&	0.16	&	68.32	&	0.045	&	14.69	&	5.61	&	0.31	&    $-$171.87	&	0.055	&	1.46	&	0.62	\\

{43 GHz:} & & & & & & & & & & \\
C	&	1.67	&	97.82	&	1.67	&	48.88	&	---	&	---	&	---	&	---	&	0.02	&	0.001	\\
J1	&	1.46	&	86.82	&	1.41	&	42.67	&	0.24	&	0.002	&    $-$176.24	&	0.01	&	0.12	&	0.005	\\
J2	&	0.06	&	24.29	&	0.06	&	12.34	&	1.6	&	0.009	&	174.6	&	0.006	&	0.07	&	0.018	\\
J3	&	0.1	&	35.36	&	0.069	&	14.46	&	2.22	&	0.051	&    $-$170.27	&	0.023	&	0.38	&	0.102	\\
J4	&	0.044	&	23.45	&	0.027	&	8.91	&	3.52	&	0.11	&    $-$175.28	&	0.031	&	0.51	&	0.22	\\
J5	&	0.051	&	23.72	&	0.026	&	8.09	&	4.62	&	0.12	&    $-$171.11	&	0.026	&	0.61	&	0.24	\\
J6	&	0.028	&	12.29	&	0.02	&	5.16	&	6.1	&	0.07	&    $-$165.27	&	0.011	&	0.4	&	0.13	\\
\midrule  
{\sc BL Lac 2015} & & & & & & & & & & \\
22 GHz: &&&&&&&&&& \\ 
C	&	1.45	&	62.53	&	1.45	&	31.27	&	---	&	---	&	---	&	---	&	0.07	&	0.002	\\
J1	&	0.21	&	44.56	&	0.21	&	22.35	&	0.81	&	0.007	&    $-$173.93	&	0.009	&	0.1	&	0.014	\\
J2	&	0.19	&	77.34	&	0.19	&	39.12	&	2.04	&	0.021	&    $-$172.4	&	0.01	&	0.16	&	0.042	\\
J3	&	0.077	&	14.21	&	0.065	&	6.49	&	3.55	&	0.031	&    $-$176.61	&	0.009	&	0.48	&	0.062	\\
J4	&	0.083	&	34.4	&	0.028	&	8.75	&	6.23	&	0.31	&    $-$171.72	&	0.049	&	1.54	&	0.62	\\
43 GHz: & & & & & & & & & & \\	
C	&	1.18	&	103.94	&	1.17	&	51.8	&	---	&	---	&	---	&	---	&	0.03	&	0.002	\\
J1	&	0.5	&	53.04	&	0.49	&	26.23	&	0.42	&	0.003	&     $-$172.11	&	0.008	&	0.1	&	0.007	\\
J2	&	0.05	&	13.8	&	0.03	&	5.35	&	1.3	&	0.051	&     $-$175.11	&	0.039	&	0.47	&	0.102	\\
J3	&	0.14	&	14.7	&	0.11	&	6.28	&	2.07	&	0.015	&     $-$171.76	&	0.007	&	0.38	&	0.029	\\
J4	&	0.043	&	12.37	&	0.023	&	4.31	&	3.14	&	0.072	&     $-$176.15	&	0.023	&	0.6	&	0.14	\\
J5	&	0.056	&	27.32	&	0.009	&	3.71	&	6.7	&	0.51	&     $-$171.87	&	0.076	&	1.89	&	1.02	\\
\bottomrule
\end{tabular}
\end{table*}

\begin{table*}[t!]
\setcounter{table}{3}
\caption{\itshape Continued}
\centering
\begin{tabular}{lcccccccccc}
\toprule
Source & $S_{Tot}$ & $\sigma_{Tot}$ & $S_{Peak}$ & $\sigma_{Peak}$ & $r$ & $\sigma_{r}$ & $\theta$ & $\sigma_{\theta}$ & $d$ & $\sigma_{d}$  \\
ID    & (Jy) & (mJy) & (Jy/beam) & (mJy/beam) & (pc) & (pc) & ($\deg$) & ($\deg$) & (pc) & (pc) \\
\midrule
{\sc 3C~111 2014} & & & & & & & & & & \\
22 GHz:&&&&&&&&&& \\ 
C	&	1.01	&	38.39	&	1.02	&	19.28	&	---	&	---	&	---	&	---	&	0	&	0	\\
J1	&	0.91	&	28.64	&	0.91	&	14.33	&	0.48	&	0.001	&	62.38	&	0.001	&	0.08	&	0.001	\\
J2	&	0.54	&	45.17	&	0.53	&	22.35	&	1.66	&	0.004	&	63.48	&	0.003	&	0.21	&	0.009	\\
J3	&	0.11	&	13.84	&	0.092	&	6.21	&	3.39	&	0.02	&	67.09	&	0.006	&	0.62	&	0.04	\\
J4	&	0.055	&	21.51	&	0.026	&	6.83	&	7.27	&	0.18	&	62.48	&	0.024	&	1.41	&	0.35	\\

43 GHz: & & & & & & & & & & \\
C	&	0.85	&	109.17	&	0.84	&	54.4	&	---	&	---	&	---	&	---	&	0.05	&	0.003	\\
J1	&	0.47	&	83.05	&	0.46	&	40.98	&	0.34	&	0.004	&	63.12	&	0.012	&	0.1	&	0.008	\\
J2	&	0.28	&	38.91	&	0.26	&	18.71	&	0.87	&	0.006	&	63.54	&	0.007	&	0.19	&	0.013	\\
J3	&	0.06	&	18.73	&	0.038	&	7.19	&	2.59	&	0.048	&	66.5	&	0.019	&	0.54	&	0.1	\\
J4	&	0.021	&	9.55	&	0.007	&	2.43	&	6.95	&	0.16	&	62.06	&	0.023	&	1	&	0.32	\\
\midrule 
{\sc 3C~111 2015} & & & & & & & & & & \\
22 GHz: &&&&&&&&& \\ 
C	&	0.52	&	26.11	&	0.52	&	13.07	&	---	&	---	&	---	&	---	&	0.03	&	0.001	\\
J0	&	0.67	&	31.43	&	0.66	&	15.7	&	0.31	&	0.001	&	57.2	&	0.002	&	0.07	&	0.002	\\
J1	&	0.44	&	54	&	0.44	&	26.97	&	1.21	&	0.003	&	61.47	&	0.002	&	0.1	&	0.006	\\
J2	&	0.21	&	15.24	&	0.21	&	7.59	&	2.3	&	0.002	&	63.53	&	0.001	&	0.1	&	0.004	\\
J3	&	0.06	&	14.86	&	0.06	&	7.23	&	4.14	&	0.014	&	65.88	&	0.003	&	0.25	&	0.028	\\
J4	&	0.013	&	3.59	&	0.012	&	1.74	&	7.95	&	0.024	&	61.16	&	0.003	&	0.34	&	0.047	\\

43 GHz: & & & & & & & & & & \\
C	&	0.98	&	114.19	&	0.97	&	56.86	&	---	&	---	&	---	&	---	&	0.05	&	0.003	\\
J0	&	0.17	&	36.19	&	0.17	&	17.78	&	0.39	&	0.005	&	61.76	&	0.013	&	0.1	&	0.01	\\
J1	&	0.21	&	52.42	&	0.16	&	22.61	&	1.19	&	0.021	&	62.27	&	0.018	&	0.32	&	0.042	\\
J2	&	0.1	&	18.66	&	0.081	&	8.36	&	2.3	&	0.014	&	62.54	&	0.006	&	0.28	&	0.028	\\
J3	&	0.027	&	9.84	&	0.028	&	5	&	4.16	&	0.003	&	65.89	&	0.001	&	0.03	&	0.006	\\
\midrule
{\sc 3C~345} & & & & & & & & & & \\
22 GHz: &&&&&&&&&& \\ 
C	&	1.94	&	164.36	&	1.9	&	81.38	&	---	&	---	&	---	&	---	&	0.15	&	0.042	\\
J1	&	0.43	&	53.42	&	0.36	&	24.46	&	8.57	&	0.11	&     $-$105.69	&	0.012	&	0.47	&	0.21	\\
J2	&	0.13	&	28.42	&	0.13	&	14.28	&	14.63	&	0.008	&     $-$89.08	&	0.001	&	0.02	&	0.015	\\
J3	&	0.28	&	110.16	&	0.08	&	24.82	&	35.4	&	1.58	&     $-$87.98	&	0.045	&	1.54	&	3.15	\\

43 GHz: & & & & & & & & & & \\
C	&	1.53	&	58.74	&	1.53	&	29.39	&	---	&	---	&	---	&	---	&	0.04	&	0.005	\\
J1	&	0.17	&	27.41	&	0.16	&	13.07	&	9.54	&	0.06	&     $-$106.38	&	0.006	&	0.22	&	0.12	\\
J0	&	0.21	&	48.78	&	0.21	&	23.98	&	2.26	&	0.043	&     $-$90.35	&	0.019	&	0.11	&	0.087	\\
J2	&	0.1	&	30.56	&	0.06	&	10.93	&	14.31	&	0.38	&     $-$89.01	&	0.027	&	0.59	&	0.76	\\
J3	&	0.052	&	32.05	&	0.02	&	8.72	&	36.26	&	1.69	&     $-$88.69	&	0.047	&	1.15	&	3.38	\\
\bottomrule
\end{tabular}
\end{table*}

\section{Conclusions \label{sec:sum}}

We presented first results from KaVA observations of eight radio-bright AGN conducted in the frame of our \emph{P}lasma-physics of \emph{A}ctive \emph{Ga}lactic \emph{N}uclei (PAGaN) project. Our main results are:

\begin{enumerate}

\item We obtained interferometric maps sufficient to identify individual source components, with numbers of components (core plus jet components) ranging from three (for $1055+015$ and $0133+476$) to seven (for BL Lac at 43 GHz).

\item We observed significant proper motions of jet components in BL Lac and 3C~111. In both sources, we found superluminal apparent jet speeds up to about about $5c$, thus constraining the intrinsic speeds to $\beta\gtrsim0.98c$ and the jet viewing angles to $\theta\lesssim20\deg$.

\item The spectral index maps for BL Lac, 3C~111, and 3C~345 show a tendency for an increasingly fast steepening of the spectral index as function of core distance in the order BL Lac object $\to$ radio galaxy $\to$ quasar. This is probably caused by systematic differences in viewing angle.

\item Jet components show systematically larger diameters $d$ at larger core distances $r$. Notably, the components of our targets all follow the same universal relation $d\approx0.2r$, albeit within substantial scatter.

\end{enumerate}

Our results demonstrate the value of KaVA for studies of the physics of active galactic nuclei. Even though our observations already provided a wealth of information, further observations will be necessary. Additional quasi-simultaneous dual-frequency observations will uncover temporal variability of spectral index maps due to jet propagation. Additional component position measurements will be important for constraining the speeds and possible non-linear motions of jet components. Overall, we may expect additional discoveries from a systematic plasma-physical study of AGN -- as intended by the PAGaN project.

\acknowledgments

This project has been supported by the Korean National Research Foundation (NRF) via Basic Research Grant 2012R1A1A2041387.
Our research has made use of data from the MOJAVE database that is maintained by the MOJAVE team \citep{Lister2009a}.
We are grateful to all staff members in KVN who helped to operate the array and to correlate the data. The KVN is a facility operated by the Korea Astronomy and Space Science Institute. The KVN operations are supported by KREONET (Korea Research Environment Open NETwork) which is managed and operated by KISTI (Korea Institute of Science and Technology Information).

\end{document}